%% file: main.tex
\documentclass[nofootinbib,a4paper,reprint,aip,longbibliography,superscriptaddress]{revtex4-1}
\usepackage[utf8]{inputenc}
\usepackage{amsmath}
\usepackage{amssymb}
\usepackage{cancel}
\usepackage{stmaryrd}
\usepackage{amsfonts}
\usepackage{bm}
\usepackage{bbm}
\usepackage{color}
\usepackage{graphicx}
\usepackage{subcaption}
\usepackage{booktabs}
\DeclareSymbolFont{operators}{OT1}{cmr}{m}{n}
\DeclareSymbolFont{letters}{OML}{cmm}{m}{it}
\DeclareSymbolFont{symbols}{OMS}{cmsy}{m}{n}
\DeclareSymbolFont{largesymbols}{OMX}{cmex}{m}{n}
\usepackage[hyperindex,breaklinks]{hyperref}
\hypersetup{
    pdfborder={0 0 0},
    allbordercolors={0 0 0},         
    unicode=false,          
    pdftoolbar=true,        
    pdfmenubar=true,        
    pdffitwindow=false,     
    pdfstartview={FitH},    
    pdfcreator={Eero Hirvijoki},
    pdfproducer={pdflatex},
    pdfnewwindow=true,
    colorlinks=true,
    linkcolor=blue,
    citecolor=blue,
    filecolor=blue,
    urlcolor=blue,  
}
\usepackage{array}
\usepackage{amsthm}
\usepackage{amsmath}
\usepackage{changes}

\newcommand{\figwidth}{0.97} 

\newcommand{\grad}{\ensuremath{\text{grad}}}
\newcommand{\curl}{\ensuremath{\text{curl}}}
\renewcommand{\div}{\ensuremath{\text{div}}}

\makeatother

\begin{document}

\title{Subcycling of particle orbits in variational, geometric electromagnetic particle-in-cell methods}

\author{Eero Hirvijoki}
\affiliation{Department of Applied Physics, Aalto University, P.O. Box 11100, 00076 AALTO, Finland}
\email{eero.hirvijoki@gmail.com}
\author{Katharina Kormann}
\affiliation{Max-Planck-Institut f\"ur Plasmaphysik, Boltzmannstra{\ss}e 2, 85748 Garching, Germany}
\affiliation{Technische Universit\"at M\"unchen, Department of Mathematics, Boltzmannstra{\ss}e 3, 85748 Garching, Germany}
\author{Filippo Zonta}
\affiliation{Department of Applied Physics, Aalto University, P.O. Box 11100, 00076 AALTO, Finland}

\date{\today}

\begin{abstract}
This paper investigates subcycling of particle orbits in variational, geometric particle-in-cell methods addressing the Vlasov--Maxwell system in magnetized plasmas. The purpose of subcycling is to allow different time steps for different particle species and, ideally, time steps longer than the electron gyroperiod for the global field solves while sampling the local cyclotron orbits accurately. The considered algorithms retain the electromagnetic gauge invariance of the discrete action, guaranteeing a local charge conservation law, while the variational approach provides a bounded long-time energy behavior. 
\end{abstract}

\maketitle 

\input{introduction.tex}
\input{basic-elements.tex}

\input{explicit-scheme.tex}
\input{explicit_push.tex}
\input{implicit-scheme.tex}

\input{hpc.tex}
\input{summary.tex}

\section*{Data availability statement}
The data that support the findings of this study has been generated with the \verb|SeLaLib| computer library \cite{selalib}. 

\begin{acknowledgments}
Financial support for the work of E.H. and F.Z. was provided by the Academy of Finland grant Nos. 315278 and 320058. Work of K.K. has been carried out within the framework of the EUROfusion Consortium and has received funding from the Euratom research and training programme 2014--2018 and 2019--2020 under grant agreement No 633053. The views and opinions expressed herein do not necessarily reflect those of the European Commission, Academy of Finland, or Aalto University.
\end{acknowledgments}


\bibliography{references}  
\end{document}

%% file: introduction.tex
\section{Introduction}
During the past decade, both understanding and developing of structure-preserving algorithms for simulating plasmas have leaped forward and, to a large extent, this development has been driven by the so-called geometric particle-in-cell (GEMPIC) methods \cite{Squire-Qin-Tang-PIC:2012PhPl,Evstatiev-shadwick:2013JCoPh,Shadwick-Stamm-Evstatiev:2014PhPl,Stamm-Shadwick-Evstatiev:2014ITPS,Xiao-et-al-kinetic:2015PhPl,He-et-al-Hamiltonian-splitting:2015PhPl,Qin-et-al:2016NucFu,Xiao-et-al-fluid:2016PhPl,Kraus-et-al:2017JPlPh,Xiao-et-al:2018PlST,Xiao-Qin-6d-tokamak:2020arXiv}---see \cite{Morrison-review:2017PhPl} for a review of the broader topic and an exhaustive list of references on the mathematical structures in plasma models. Based on discretizing either the underlying variational or Hamiltonian structure, GEMPIC algorithms provide long-time fidelity and stability for models with possibly billions of degrees of freedom. This is especially important for kinetic simulations of magnetized fusion plasmas where reaching macroscopic transport at time scales of $10^{-6} \text{s}$ requires a breathtaking number of time steps to resolve the electron cyclotron motion typically appearing at the time scales of  $10^{-11} \text{s}$. Such an enormous feat has been performed only very recently \cite{Xiao-Qin-6d-tokamak:2020arXiv} but this will likely become common place during 2020s. 

To our knowledge, the GEMPIC methods have so far considered only synchronous integration of particle orbits and electromagnetic fields, whereas the non-GEMPIC methods, that have become the industry standard \cite{Adam-et-al:1982JCoPh,COHEN1985311,IPIC3D:2010,Chen:2011jf,Chacon:2013cz,Chen:2014eh,IPIC3D_subcycling:2015,Chacon:2016gi,Chen-et-al:2019arXiv190301565C}, implement so-called subcycling or orbit-averaging of particle orbits out of the box. The only GEMPIC attempt in this direction,  reported in \cite{Kormann2019}, is based on an energy-conserving temporal discretization rather than a variational integrator. Especially in simulating multi-component, strongly magnetized plasmas, treating both ions and electrons kinetically, multiple different time scales naturally emerge as the ion and electron cyclotron periods differ by the respective mass ratio. It would be preferable not to restrict the field solve or the ion push to the fastest time scale---typically the electron cyclotron period unless the plasma density is very high, in which case the electron plasma oscillations become dominant---but to allow for subcycling of particle orbits at their naturally occurring frequencies. The absence of this feature from the GEMPIC methods does not need to remain the state of the business, though, and the purpose of this paper is to explore the possibilities in modifying GEMPIC methods towards fully asynchronous and, in future, possibly temporally adaptive integration. 

We investigate two different strategies for subcycling of particle orbits within the variational GEMPIC framework. Both approaches retain the electromagnetic gauge invariance of the discrete action---guaranteeing a local charge conservation law---and the variational approach provides a bounded long-time energy behavior. The first approach is intended for upgrading the existing variational GEMPIC methods to include subcycling with minimal effort invested in modifications: the global field solves are explicit and the local particle push implicit for each particle individually, just as in the pioneering paper \cite{Squire-Qin-Tang-PIC:2012PhPl}, or explicit if rectilinear meshes with a diagonal metric and the novel zigzag path~\cite{Xiao-et-al:2018PlST,Xiao-Qin-6d-tokamak:2020arXiv} are exploited. The requirement for gauge invariance, however, leads to a peculiarity: the magnetic field is orbit-averaged but the effect of electric field on the particle orbits is evaluated only once during the sybcycling period. Numerical tests indicate that artificial oscillations may occur if the electric field impulse on the particle orbit is too large, essentially when the global time step approaches or exceeds the cyclotron time scale. 
This behavior is likely credited to the electric field not being orbit-averaged the same way as the magnetic field is which increases the instantaneous relative impulse from the electric field in comparison to the impulse from the magnetic field. Indeed, the oscillations are observed to vanish if orbit-averaging is enforced also for the electric field but then the particle push is no longer variational and the good long-time behaviour is destroyed. 
    
The second strategy is proposed to remedy the issues possibly occurring with the first approach. Instead of relying on the ``summation-by-parts'' trick, we observe that, in enabling proper partial integration in the field--particle-interaction term of the discrete action, both magnetic and electric field impulses can be orbit-averaged and the variational structure together with gauge invariance retained. Repeating the numerical tests confirms our hypothesis that the artificial oscillations are rooted in not orbit-averaging the electric field impulse, since the second scheme does not exhibit such artificial oscillations. The choice of enabling proper partial integration in the discrete action, however, appears to always lead to an implicit scheme, in contrast to the clever summation-by-parts trick that admits explicit field solve and a fully explicit scheme in case of rectilinear meshes. This implicitness, however, only relates to how the electric field, the current density, and the particle push are coupled for the Faraday equation remains explicit. Consequently, both algorithms have CFL-conditions on the field solves. Per these findings, it remains to be seen if proper orbit-averaging could be performed within the variational framework with explicit schemes. 


We will begin by briefly recapitulating the essential elements of a structure-preserving variational discretization of the Vlasov--Maxwell system in Sec.~\ref{sec:basics}, and then proceed to presenting the new algorithms. The explicit scheme with implicit particle push is introduced in Sec.~\ref{sec:explicit} together with the numerical experiments indicating the possible oscillation problem and a demonstration that brute-force orbit-averaging the electric field removes them. The version with explicit particle push, requiring rectilinear meshes is provided in Sec.~\ref{sec:explicit-push}, demonstrating similar behavior. Building on this learning outcome, the implicit scheme is derived in Sec.~\ref{sec:implicit} and the numerical tests repeated yet again, demonstrating that the artificial oscillations no longer exist and that the global step size may safely exceed the cyclotron period as long as it remains within the CFL limit. Finally, we engage in a brief discussion regarding the high-performance-computing aspects and stability of the algorithms in Sec.~\ref{sec:hpc} while the results are summarized and possible suggestions for future research directions discussed in Sec.~\ref{sec:summary}. 



%% file: basic-elements.tex
\section{Elements of structure-preserving discretization}\label{sec:basics}
In this section, we briefly summarize some of the essential building blocks for implementing a variational GEMPIC method for the Vlasov--Maxwell system. For more details, we refer the reader to the excellent papers \cite{Squire-Qin-Tang-PIC:2012PhPl,Kraus-et-al:2017JPlPh,Xiao-et-al:2018PlST}.

Let us assume we have some domain $\Omega\subset\mathbb{R}^3$ and a finite-dimensional discretization of the associated de Rham complex: we expect there to be the sets of scalar and vector valued basis functions $\{W^0_i\}_i$, $\{\bm{W}^1_j\}_j$, $\{\bm{W}^2_k\}_k$, and $\{W^3_\ell\}_\ell$, all functions of position $\bm{x}$, such that
\begin{align}
    \nabla W^0_i(\bm{x})&=\grad_i^j\bm{W}^1_j(\bm{x}),\\
    \nabla\times\bm{W}^1_j(\bm{x})&=\curl_j^k\bm{W}^2_k(\bm{x}),\\
    \nabla\cdot\bm{W}^2_k(\bm{x})&=\div_k^{\ell}W^3_{\ell}(\bm{x}),
\end{align}
where $\grad_i^j$, $\curl_j^k$, and $\div_k^{\ell}$ denote the elements of the discrete gradient, curl, and divergence matrices, respectively. Einstein summation over the repeated superscript-subscript index pairs is assumed throughout, and the letters $i,j,k,\ell$ always refer to the corresponding element spaces as denoted above. One typical way to construct such basis is via the Whitney interpolating functions on simplical meshes while structured rectilinear meshes often use regular polynomials.

Because the basis functions satisfy the de Rham complex, we have that
\begin{align}
    0&=\nabla\times\nabla W^0_i=\grad_i^j\nabla\times\bm{W}^1_j=\grad_i^j\curl_j^k \bm{W}^2_k,\\
    0&=\nabla\cdot\nabla\times\bm{W}^1_j=\curl_j^k\nabla\cdot\bm{W}^2_k=\curl_j^k\div_k^\ell W^3_\ell,
\end{align}
for all $\bm{x}\in\Omega$, implying the matrix identities $\curl_j^k\grad_i^j=0$ and $\div_k^\ell\curl_j^k=0$. The spatial discretizations of the vector and scalar potential are then taken to be
\begin{align}
    \bm{A}_{\text{ext}}(\bm{x})&=a_{\text{ext}}^j\bm{W}^1_j(\bm{x}),\\
    \bm{A}(\bm{x},t)&=a^j(t)\bm{W}^1_j(\bm{x}),\\
    \phi(\bm{x},t)&=\phi^i(t)W^0_i(\bm{x}),
\end{align}
where the subscript ``$\text{ext}$'' refers to a static, given quantity, and the definitions then imply the following expressions for the finite-dimensional electric and magnetic fields
\begin{align}
    \bm{E}&=(-\dot{a}^j-\phi^i\grad_i^j)\bm{W}^1_j\equiv e^j\bm{W}^1_j,\\
    \bm{B}&=a^j\curl_j^k\bm{W}^2_k\equiv b^k\bm{W}^2_k.
\end{align}
A possible external, fixed magnetic field is naturally denoted by 
\begin{align}
    \bm{B}_{\text{ext}}&=a_{\text{ext}}^j\curl_j^k\bm{W}^2_k\equiv b_{\text{ext}}^k\bm{W}^2_k.
\end{align}
The finite-dimensional magnetic field now satisfies the identities
\begin{align}
    \partial_t\bm{B}&=\dot{a}^j\curl_j^k\bm{W}^2_k=-e^j\nabla\times\bm{W}^1_j=-\nabla\times\bm{E},
    \\
    \partial_t\nabla\cdot\bm{B}&=\dot{a}^j\curl^k_j\div_k^\ell W^3_\ell=0,
\end{align}
meaning that, if the degrees of freedom for $\bm{B}$ initially satisfy $b^k\div^{\ell}_k=0$, the magnetic field will stay divergence free for all times.

In particle-in-cell methods, the idea is to let marker particles to carry the phase-space density forward in time, starting from a fixed initial density distribution 
\begin{align}
    F_0=\sum_p\delta(\bm{x}_0-\bm{x}_p(t_0))\delta(\bm{v}_0-\dot{\bm{x}}_p(t_0)),
\end{align}
where $(\bm{x}_p(t_0),\dot{\bm{x}}_p(t_0))$ are the initial position and velocity coordinates for the marker trajectory $(\bm{x}_p(t),\dot{\bm{x}}_p(t))$. In practice, every marker should be weighted with a label $w_p$ accounting for the number of real particles the marker represents. Here we have, however, suppressed this factor for notational clarity. From here on, we will also use the tuples $\mathbbm{x}=\{\bm{x}_p\}_p$, $\dot{\mathbbm{x}}=\{\dot{\bm{x}}_p\}_p$, $\mathbbm{a}=\{a^j\}_j$, $\dot{\mathbbm{a}}=\{\dot{a}^j\}_j$ $\mathbbm{b}=\{b^k\}_k$, $\mathbbm{e}=\{e^j\}_j$, and $\phi=\{\phi^i\}_i$ to group together the degrees of freedom. Especially it is to be understood that $\phi$ now refers to the tuple of degrees of freedom, not to the space-continuous electrostatic potential.  

Once the above definitions are cleared, one substitutes them to the Vlasov--Maxwell action functional, performs the integrations over phase space, and obtains a finite-dimensional yet time-continuous action functional
\begin{align}
    S[\mathbbm{x},\mathbbm{a},\phi]&=\int_{t_\text{i}}^{t_\text{f}}\frac{\varepsilon_0}{2}e^{j_1}M^1_{j_1,j_2}e^{j_2}dt\nonumber
    \\
    &-\int_{t_\text{i}}^{t_\text{f}}\frac{\mu_0^{-1}}{2}(b^{k_1}+b_{\text{ext}}^{k_1})M^2_{k_1k_2}(b^{k_2}+b_{\text{ext}}^{k_2})dt
    \nonumber
    \\
    &+\sum_p \int_{t_\text{i}}^{t_\text{f}}q (a^j+a_{\text{ext}}^j)\bm{W}^1_j(\bm{x}_p)\cdot\dot{\bm{x}}_pdt
    \nonumber
    \\
    &-\sum_p\int_{t_\text{i}}^{t_\text{f}}q\phi^iW_i^0(\bm{x}_p)dt
    \nonumber
    \\
    &+\sum_p\int_{t_\text{i}}^{t_\text{f}}\frac{1}{2}m|\dot{\bm{x}}_p|^2dt,
\end{align}
where one is to remember the relations $e^j=-\dot{a}^j-\grad_i^j\phi^i$ and $b^k=\curl_j^ka^j$. The constant finite-element mass matrices, related to one-form and two-form element bases, are defined according to
\begin{align}
    \int_\Omega \bm{W}^1_{j_1}(\bm{x})\cdot\bm{W}^1_{j_2}(\bm{x})d\bm{x}=M^1_{j_1j_2},\\
    \int_\Omega \bm{W}^2_{k_1}(\bm{x})\cdot\bm{W}^2_{k_2}(\bm{x})d\bm{x}=M^2_{k_1k_2}.
\end{align}

From the perspectives of solving the Vlasov--Maxwell system of equations while respecting the Gauss' law constraints, the electromagnetic gauge invariance turns out to be a key requirement. Let us first perturb $\mathbbm{a}\rightarrow\mathbbm{a}+\epsilon\delta \mathbbm{a}$ and $\phi\rightarrow\phi+\epsilon\delta\phi$ and differentiate the perturbed action with respect to $\epsilon$ at $\epsilon=0$. This computation provides
\begin{align}
    &\partial_{\epsilon}|_{\epsilon=0}S[\mathbbm{x},\mathbbm{a}+\epsilon\delta\mathbbm{a},\phi+\epsilon\delta\phi]\nonumber
    \\
    &=-\int_{t_\text{i}}^{t_\text{f}}\frac{d}{dt}\left(\delta a^{j_1}\varepsilon_0M^1_{j_1,j_2}e^{j_2}\right)dt
    \nonumber\\
    &+\int_{t_\text{i}}^{t_\text{f}}\delta a^{j_1}\left(\varepsilon_0M^1_{j_1,j_2}\dot{e}^{j_2}-\mu_0^{-1}\curl_{j_1}^{k_1}M^2_{k_1,k_2}(b^{k_2}+b^{k_2}_{\text{ext}})\right)dt\nonumber
    \\
    &-\int_{t_\text{i}}^{t_\text{f}}\delta\phi^{i_1}\varepsilon_0\grad_{i_1}^{j_1}M^1_{j_1,j_2}e^{j_2}dt\nonumber
    \\
    &+\sum_p \int_{t_\text{i}}^{t_\text{f}}\left( q\delta a^j\bm{W}_j^1(\bm{x}_p)\cdot\dot{\bm{x}}_p-q\delta \phi^iW^0_i(\bm{x}_p)\right) dt.
\end{align}
Applying Hamilton's principle of least action, assuming the perturbations $\delta \mathbbm{a}$ and $\delta\phi$ to be arbitrary and to vanish at $t_\text{i}$ and $t_\text{f}$, the Euler--Lagrange equations correspond to a discrete Amp\`ere--Maxwell equation
\begin{align}
    \label{eq:time-continuous-space-discrete-ampere}    
    &\varepsilon_0M^1_{j_1,j_2}\dot{e}^{j_2}+J_j
    =\mu_0^{-1}\curl_{j_1}^{k_1}M^2_{k_1,k_2}(b^{k_2}+b^{k_2}_{\text{ext}}),
\end{align} 
and a discrete Gauss's law for the electric field
\begin{align}
    \label{eq:time-continuous-space-discrete-Gauss}
    -\varepsilon_0\grad_{i_1}^{j_1}M^1_{j_1,j_2}e^{j_2}&=\varrho_i,
\end{align}
with current density $J_j=\sum_pq\bm{W}^1_j(\bm{x}_p)\cdot\dot{\bm{x}}_p$ and charge density $\varrho_i=\sum_pqW^0_i(\bm{x}_p)$. If, however, we choose the very specific forms for the perturbations
\begin{align}
    \delta a^j&=\chi^i\grad_i^j,\\
    \delta\phi^i&=-\dot{\chi}^i,
\end{align}
requesting that $\chi^i(t_\text{i})=\chi^i(t_\text{f})=0$, we observe that the differentiation of the transformed action can now be written as
\begin{align}
    &\partial_{\epsilon}|_{\epsilon=0}S[\mathbbm{x},a^j+\epsilon\chi^i\grad_i^j,\phi^i-\epsilon\dot{\chi}^i]\nonumber
    \\
    &=\int_{t_\text{i}}^{t_\text{f}}\chi^i\left(\grad_i^jJ_j-\dot{\varrho}_i\right)dt.
\end{align}
Because the action also has a strong symmetry with respect to arbitrary $\chi^i$ in the sense that 
\begin{align}
    S[\mathbbm{x},a^j+\epsilon\chi^i\grad_i^j,\phi^i-\epsilon\dot{\chi}^i]=S[\mathbbm{x},\mathbbm{a},\phi],
\end{align}
the differentiation of the transformed action with respect to $\epsilon$ has to vanish, providing the finite-dimensional charge conservation law
\begin{align}
    \grad_i^jJ_{j}-\dot{\varrho}_i=0.
\end{align}
The importance of this identity lies in the fact that it eliminates the need to solve the Gauss' law: solving for the electric-field $e^j(t)$ in \eqref{eq:time-continuous-space-discrete-ampere} guarantees such evolution for $e^j(t)$ that it automatically satisfies the Gauss' law \eqref{eq:time-continuous-space-discrete-Gauss}. It is then a matter of finding a temporal discretization which retains an analog of this property also in the fully discrete case. 


%% file: explicit-scheme.tex
\section{Subcycling of particles with an explicit field solve}
\label{sec:explicit}
Turning into the details of implementing a subcycling scheme, we first investigate a straigthforward modification of the pioneering scheme~\cite{Squire-Qin-Tang-PIC:2012PhPl}. We obtain an algorithm where the particle push is implicit and the field solve explicit, just as in the pioneering work---an explicit particle push, exploiting the zigzag path and rectilinear meshes, is discussed in Sec.~\ref{sec:explicit-push}. Requesting electromagnetic gauge invariance, the subcycling turns out such that the magnetic field is properly orbit-averaged but the effect of the electric field on the particle orbits is evaluated only once during the sybcycling period. Numerical tests then suggest that, if the global time step is too long, the resulting large impulse from a single electric kick might lead to artificial oscillations. 
Enforcing orbit-averaging also for the electric field removes the artificial oscillations but renders the algorithm to no longer be variational. 

\subsection{Discrete equations}
The time integral in the action functional is split into intervals $[t_n,t_{n+1}]$, here of equal length $\Delta t$, and the total action then comprises of the sum
\begin{align}
    S
    =\sum_{n=0}^{N-1}S_{n,n+1} .
\end{align}
One then assumes some discrete representations for the variable paths in the intervals $t\in [t_n,t_{n+1}]$ and approximates the $S_{n,n+1}$, typically with some quadrature rule. 
Here, we closely follow the pioneering work \cite{Squire-Qin-Tang-PIC:2012PhPl} but introduce a modification, allowing for subcycling of the particles with $\mathcal{V}$ indicating the number of substeps per global time step $\Delta t$. 

The discrete action over the time interval $t\in[t_n,t_{n+1}]$ we approximate with the expression
\begin{align}\label{eq:discrete-action-k-multisteps}
    &S_{n,n+1}[\mathbbm{x}_n, \mathbbm{x}_{n+1/\mathcal{V}}, \mathbbm{x}_{n+2/\mathcal{V}},\ldots ,\mathbbm{x}_{n+1},\mathbbm{a}_n,\mathbbm{a}_{n+1},\phi_n]\nonumber\\
    &=\Delta t\frac{\varepsilon_0}{2}e^{j_1}_nM^1_{j_1,j_2}e^{j_2}_n\nonumber\\
    &-\Delta t\frac{\mu_0^{-1}}{2}(b_n^{k_1}+b_{\text{ext}}^{k_1})M^2_{k_1k_2}(b_n^{k_2}+b_{\text{ext}}^{k_2})
    \nonumber
    \\
    &+\sum_{p} \sum_{\nu=1}^{\mathcal{V}} q (a_{n+1}^j+a_{\text{ext}}^j)\int_{0}^{1}\bm{W}^1_j(\bm{x}_{p}^{n,\nu}(\tau))\cdot\frac{d\bm{x}_{p}^{n,\nu}(\tau)}{d\tau}d\tau
    \nonumber
    \\
    &-\sum_pq\phi_n^iW_i^0(\bm{x}_{p,n})\Delta t
    \nonumber
    \\
    &+\sum_p \sum_{\nu=1}^{\mathcal{V}} \frac{m}{2}\frac{|\bm{x}_{p,n+\nu/\mathcal{V}}-\bm{x}_{p,n+(\nu - 1)/\mathcal{V}}|^2}{\Delta t/\mathcal{V}},
\end{align}
where the following abbreviations have been introduced
\begin{align}\label{eq:electric-magnetic-field_multistep}
    b^k_n&=a^j_n\curl_j^k,\\
    e^j_n&=-(a^j_{n+1}-a^j_{n})/\Delta t-\phi^i_{n}\grad^j_i,
\end{align}
and $\bm{x}_{p}^{n,\nu}(\tau)$ is a straight trajectory connecting the substeps $(\nu - 1)$ and $\nu$ linked to a global step $n$ and defined according to
\begin{align}
    \bm{x}_{p}^{n,\nu}(\tau)
    &=(1-\tau)\bm{x}_{p,n+(\nu-1)/\mathcal{V}}+\tau \bm{x}_{p,n+\nu/\mathcal{V}}.
\end{align}

The discrete Euler--Lagrange conditions are derived by perturbing the variables, assuming the perturbations to vanish at the end points in time, and looking for an extrema point of the discrete action. With respect to the perturbations $\mathbbm{a}_n\rightarrow \mathbbm{a}_n+\epsilon\delta \mathbbm{a}_n$, this leads to the equation
\begin{align}
    &\partial_{\epsilon}|_{\epsilon=0}S_{n,n+1}[\mathbbm{a}_n+\epsilon\delta \mathbbm{a}_n]
    \nonumber\\
    &+\partial_{\epsilon}|_{\epsilon=0}S_{n-1,n}[\mathbbm{a}_{n}+\epsilon\delta \mathbbm{a}_{n}]=0,
\end{align}
and, when written explicitly, provides the discrete Amp\`ere--Maxwell equation
\begin{align}\label{eq:discrete-ampere_multistep}
    &\varepsilon_0M^1_{j,j_2}\frac{e_{n+1}^{j_2}-e^{j_2}_{n}}{\Delta t}+J^{n,n+1}_j=\mu_0^{-1}\curl_{j}^{k} M^2_{k,k_2}(b_{n+1}^{k_2}+b_{\text{ext}}^{k_2}),
\end{align}
with a discrete current density defined according to
\begin{align}\label{eq:explicit-current-density}
    J_j^{n,n+1}&=\sum_p \sum_{\nu=1}^{\mathcal{V}} q \int_{0}^{1}\bm{W}^1_j(\bm{x}_{p}^{n, \nu}(\tau))\cdot\frac{d\bm{x}_{p}^{n, \nu}(\tau)}{d\tau}\frac{d\tau}{\Delta t}.
\end{align}
With respect to perturbations $\phi_n\rightarrow\phi_n+\epsilon\delta\phi_n$, the variation of the action leads to 
\begin{align}
    \partial_{\epsilon}|_{\epsilon=0}S_{n,n+1}[\phi_n+\epsilon\delta \phi_n]=0,
\end{align}
which, when written explicitly, corresponds to the discrete Gauss' law
\begin{align}\label{eq:discrete-gauss_multistep}
    \varrho_i^n&=-\varepsilon_0\grad_i^jM^1_{j,j_2}e^{j_2}_n,
\end{align}
with the discrete charge density being defined according to
\begin{align}\label{eq:discrete-charge-density}
    \varrho_i^n&=\sum_pq W_i^0(\bm{x}_{p,n}).
\end{align}

Perturbing the particles' spatial positions $\mathbbm{x}_n\rightarrow \mathbbm{x}_n+\epsilon\delta\mathbbm{x}_n$ provides
\begin{align}
    &\partial_{\epsilon}|_{\epsilon=0}S_{n,n+1}[\mathbbm{x}_{n}+\epsilon\delta\mathbbm{x}_n]\nonumber\\
    &+\partial_{\epsilon}|_{\epsilon=0}S_{n-1,n}[\mathbbm{x}_{n}+\epsilon\delta\mathbbm{x}_n]=0,
\end{align}
while perturbing $\mathbbm{x}_{n+\nu/\mathcal{V}} \rightarrow \mathbbm{x}_{n+\nu/\mathcal{V}}+\epsilon\delta\mathbbm{x}_{n+\nu/\mathcal{V}}$ provides 
\begin{align}
    &\partial_{\epsilon}|_{\epsilon=0}S_{n,n+1}[\mathbbm{x}_{n+\nu/\mathcal{V}}+\epsilon\delta\mathbbm{x}_{n+\nu/\mathcal{V}}]=0,
\end{align}
for each $n = 0, \ldots, N - 1$ and $\nu = 1, \ldots, \mathcal{V} - 1$.
Written explicitly, these correspond to the equations for the indices $n$
\begin{align}\label{eq:discrete-el-x-integer_multistep}
    &m\frac{\bm{x}_{p,n+1/\mathcal{V}}-2\bm{x}_{p,n}+\bm{x}_{p,n-1/\mathcal{V}}}{(\Delta t/\mathcal{V})^ 2}
    \nonumber
    \\
    &=q\frac{\bm{x}_{n+1/\mathcal{V}}-\bm{x}_{p,n}}{\Delta t/\mathcal{V}}
    \nonumber\\
    &\qquad\qquad\times \int_0^1(1-\tau)(b_{n+1}^k+b_{\text{ext}}^k)\bm{W}^2_k(\bm{x}_{p}^{n,1}(\tau))d\tau
    \nonumber
    \\
    &+q\frac{\bm{x}_{p,n}-\bm{x}_{p,n-1/\mathcal{V}}}{\Delta t/\mathcal{V}}\nonumber\\
    &\qquad\qquad\times \int_0^1\tau (b_n^k+b_{\text{ext}}^k)\bm{W}^2_k(\bm{x}_{p}^{n, 0}(\tau))d\tau
    \nonumber
    \\
    &+q \mathcal{V} e_n^j\bm{W}^1_j(\bm{x}_{p,n}).
\end{align}
and for the indices $\nu$
\begin{align}\label{eq:discrete-el-x-half-integer_multistep}
    &m\frac{\bm{x}_{p,n+(\nu+1)/\mathcal{V}}-2\bm{x}_{p,n+\nu/\mathcal{V}}+\bm{x}_{p,n+(\nu-1)/\mathcal{V}}}{(\Delta t/\mathcal{V})^2}
    \nonumber
    \\
    &=q\frac{\bm{x}_{n+(\nu+1)/\mathcal{V}}-\bm{x}_{p,n+\nu/\mathcal{V}}}{\Delta t/\mathcal{V}}
    \nonumber\\
    &\qquad\qquad\times \int_0^1(1-\tau)(b_{n+1}^k+b_{\text{ext}}^k)\bm{W}^2_k(\bm{x}_{p}^{n,\nu+1}(\tau))d\tau
    \nonumber
    \\
    &+q\frac{\bm{x}_{p,n+\nu/\mathcal{V}}-\bm{x}_{p,n+(\nu-1)/\mathcal{V}}}{\Delta t/\mathcal{V}}
    \nonumber\\
    &\qquad\qquad\times \int_0^1\tau (b_{n+1}^k+b_{\text{ext}}^k)\bm{W}^2_k(\bm{x}_{p}^{n,\nu}(\tau))d\tau .
\end{align}
Note that the electric-field impulse is evaluated only for steps with index $n$, not for the $\nu$.

The equations \eqref{eq:discrete-ampere_multistep}, \eqref{eq:discrete-gauss_multistep}, \eqref{eq:discrete-el-x-integer_multistep}, and \eqref{eq:discrete-el-x-half-integer_multistep} are completed by the discrete Faraday equation that is a direct consequence of the definitions for $\mathbbm{e}_n,\mathbbm{b}_n$, namely
\begin{align}\label{eq:discrete-faraday_multistep}
    \frac{b^k_{n}-b_{n-1}^k}{\Delta t}=-\curl_j^ke_{n-1}^j.
\end{align}

The electromagnetic gauge invariance and the discrete charge conservation law are verified in the following manner. Let
\begin{align}
    &a^j_n\rightarrow a^j_n+\chi^i_n\grad^j_i, \\ &\phi^i_n\rightarrow\phi^i_n-\frac{\chi^i_{n+1}-\chi^i_{n}}{\Delta t},
\end{align}
and the total discrete action \eqref{eq:discrete-action-k-multisteps} will satisfy the strong symmetry condition
\begin{align}
    &\sum_{n=0}^{N-1}S_{n,n+1}\big[a^j_n+\chi^i_n\grad_i^j,a^j_{n+1}+\chi^i_{n+1}\grad_i^j,\nonumber
    \\
    &\qquad\qquad\qquad\phi^i_n-(\chi^i_{n+1}-\chi^i_n)/\Delta t\big]\nonumber
    \\
    &=\sum_{n=0}^{N-1}S_{n,n+1}[\mathbbm{a}_n,\mathbbm{a}_{n+1},\phi_n]\nonumber
    \\
    &+\sum_p e_p\big[\chi_{N}^iW_i^0(\bm{x}_{p,N})-\chi^i_{0}W_i^0(\bm{x}_{p,0})\big].
\end{align}
Differentiating with respect to $\chi_{n}$ at any $n$ such that $n\neq 0$ and $n\neq N$, the right side vanishes identically as it is independent of $\chi_{n}$, and one finds the discrete charge conservation law
\begin{align}
    \grad_i^jJ_{j}^{n-1,n}-\frac{\varrho^{n}_i-\varrho_i^{n-1}}{\Delta t}=0.
\end{align}
The significance of this equation is that, if we assume the Gauss' law \eqref{eq:discrete-gauss_multistep} to hold for $n-1$, the charge conservation and the Amp\`ere equation \eqref{eq:discrete-ampere_multistep} then imply
\begin{align}
    \varrho^{n}_i=&\varrho_i^{n-1}+\Delta t\,\grad_i^jJ_{j}^{n-1,n}=-\varepsilon_0\grad_i^jM^1_{j,j_2}e^{j_2}_n,
\end{align}
meaning that the Gauss' law is automatically satisfied, if it is satisfied initially. 

Together the discrete equations provide means of advancing the degrees of freedom $\mathbbm{x}_n$, $\mathbbm{e}_n$, and $\mathbbm{b}_n$ in time according to the following strategy
\begin{enumerate}
    \setcounter{enumi}{-1}
    \item Given $\mathbbm{x}_0$, initialize $\mathbbm{e}_0$ with the Gauss's law \eqref{eq:discrete-gauss_multistep} and approximate $\mathbbm{x}_{-1/\mathcal{V}}$ using $\mathbbm{v}_0$
    \item Given $\mathbbm{e}_{n},\mathbbm{b}_n$, compute $\mathbbm{b}_{n+1}$ from the Faraday equation \eqref{eq:discrete-faraday_multistep}
    \item Given $(\mathbbm{e}_n,\mathbbm{b}_n,\mathbbm{b}_{n+1})$ and $(\mathbbm{x}_{n-1/\mathcal{V}},\mathbbm{x}_{n})$, push markers with \eqref{eq:discrete-el-x-integer_multistep} to obtain $\mathbbm{x}_{n+1/\mathcal{V}}$
    \item Given $\mathbbm{x}_n,\mathbbm{x}_{n+1/\mathcal{V}}$ and $\mathbbm{b}_n,\mathbbm{b}_{n+1}$, push markers to $\mathbbm{x}_{n+2/\mathcal{V}},...,\mathbbm{x}_{n+1}$ with \eqref{eq:discrete-el-x-half-integer_multistep} and accumulate $J_{j}^{n,n+1}$ according to \eqref{eq:explicit-current-density}
    \item Given $\mathbbm{e}_{n},\mathbbm{b}_{n+1}$ and the recorded value for $J_{j}^{n,n+1}$, invert the Amp\`ere-Maxwell equation \eqref{eq:discrete-ampere_multistep} for $\mathbbm{e}_{n+1}$
    \item Repeat the steps 1-4 indefinitely
\end{enumerate}

\subsection{Numerical tests}
We have implemented the method within the GEMPIC code in the library \verb|SeLaLib| \cite{selalib}. The code is based on compatible spline-finite-element bases as described in~\cite{Kraus-et-al:2017JPlPh}. For our experiments, we have chosen a solver based on cubic and quadratic splines. Remember that without subcycling, the algorithm corresponds to the pioneering variational scheme~\cite{Squire-Qin-Tang-PIC:2012PhPl}, which will be used for verification. The scheme contains a non-linearity in equations \eqref{eq:discrete-el-x-integer_multistep} and \eqref{eq:discrete-el-x-half-integer_multistep}. This non-linearity, however, only couples the three components of the three positions of each particle. This non-linear step can efficiently be solved by a first guess obtained by extrapolation from the old values, followed by one or more updates according to Newton's method. For this, an analytic formula for the derivative matrix can be found and evaluated numerically in the implementation. In our experiments, we consider the Newton iteration to be converged at a tolerance of $10^{-10}$.

\subsubsection{An electrostatically dominated test case}

As a first example of a simulation with strong background magnetic field, we consider a reduced,  1D-2V-dimensional phase space and an initial distribution function of
\begin{equation}
    f(x,v_1,v_2) = \frac{1 + 0.1 \cos(0.5x)}{2\pi} \exp\left(-\frac{v_1^2+v_2^2}{2}\right),
\end{equation}
set up in a background magnetic field of $B_3(x,0) = 2\pi 10$. We run the variational subcycling and the Hamiltonian splitting algorithms with 32 grid points and 160,000 particles until time 20. 
In this case, the spatial resolution is given by $\Delta x = 4\pi/32$. Since we split the curl-part in Faraday's and Amp\`ere's law, we get a stability limit of $\Delta t < \Delta x\sqrt{17/42} \approx 0.2498$ (cf.~\cite[Appendix A2]{Kormann2019}). 
For our choice of the magnetic field, the cyclotron period is~0.1.

\begin{figure}
\centering
\begin{subfigure}{\figwidth\linewidth}
    \includegraphics[width=\figwidth\linewidth]{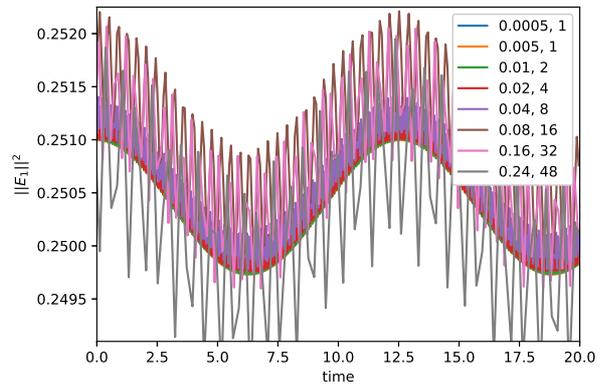}
    \caption{Subcycling ($\Delta \tau = 0.005$).}
    \label{fig:explicit_sub20_landau}
\end{subfigure}
~
\begin{subfigure}{\figwidth\linewidth}
    \includegraphics[width=\figwidth\linewidth]{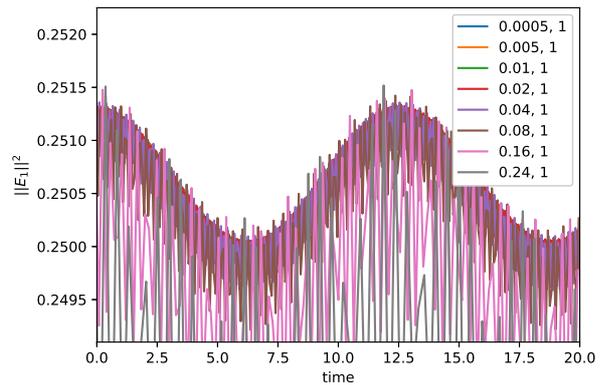}
    \caption{No subcycling.}
    \label{fig:explicit_landau}
\end{subfigure}
\caption{Electrostatically dominated test case: Time evolution of $\|E_1\|^2$ for varying global time steps (given in the legends together with the number of substeps) with $B(x,0) = 2\pi 10$. }
\end{figure}

Figure~\ref{fig:explicit_sub20_landau} shows the temporal dynamics of the first component of the electric energy with respect to an increasing global time step and a fixed substep length of $\Delta \tau = 0.005$ (20 substeps per cyclotron period). The results up to $\Delta t = 0.04$ are reasonably resolved. For $\Delta t = (0.08, 0.16, 0.32)$ we see that the solution becomes clearly less accurate as the global step approaches and exceeds the cyclotron period 0.1 but the macroscale behavior is nevertheless approximately retained. The number of Newton steps stays constant in all runs and is around 2.6 on average (including the initial step) with substeps. Without substeps it increases slightly with the time step to reach around 3.2 iterations on average for $\Delta t = 0.24$. A comparison to the standard scheme, with no subcycling is presented in Fig.~\ref{fig:explicit_landau}. One can see that the cyclotron-scale oscillations cannot be sampled anymore as in the case with substepping and, as a consequence, also the qualitative macroscale behavior is significantly worse than with substeps. In complicated magnetic fields such as those encountered in tokamaks and stellarators, we would expect the results without subcycling to look even worse: the choice of using 10--20 steps per cyclotron period is a rather common strategy and often necessary in fusion-related orbit-following applications. 
Table~\ref{tab:landauB_Gauss} confirms that the scheme is conserving Gauss' law with and without substeps, and Table~\ref{tab:landauB_energy} shows the error in energy conservation of the scheme which is rather small. 



\begin{table}
\begin{subtable}{0.48\textwidth}
\begin{tabular}{|c|c|c|}
\hline
     $\Delta t$  & with substeps & without substeps \\
     \hline 
      0.005 &    & $1.21 \cdot 10^{-14}$ \\
      0.01 &   $1.20 \cdot 10^{-14}$ & $1.21 \cdot 10^{-14}$ \\
      0.02 &   $1.20 \cdot 10^{-14}$& $1.22 \cdot 10^{-14}$ \\
      0.04 &    $1.12 \cdot 10^{-14}$ & $1.11 \cdot 10^{-14}$ \\
      0.08 &    $1.27 \cdot 10^{-14}$ & $1.69 \cdot 10^{-14}$ \\
      0.16 &  $1.10 \cdot 10^{-14}$ &$1.07 \cdot 10^{-14}$ \\
      0.24 & $1.23 \cdot 10^{-14}$ &$1.15 \cdot 10^{-14}$ \\
      \hline
\end{tabular}
\caption{Conservation of Gauss' law}
\label{tab:landauB_Gauss}
\end{subtable}
~
\begin{subtable}{0.48\textwidth}
\begin{tabular}{|c|c|c|}
\hline
     $\Delta t$&  with substeps & without substeps \\
     \hline
      0.005 &    & $7.94 \cdot 10^{-10}$ \\
      0.01 &   $1.42 \cdot 10^{-9}$ & $2.08 \cdot 10^{-9}$ \\
      0.02 &   $2.97 \cdot 10^{-9}$& $5.34 \cdot 10^{-9}$ \\
      0.04 &    $8.32 \cdot 10^{-9}$ & $1.09 \cdot 10^{-8}$ \\
      0.08 &    $9.91 \cdot 10^{-8}$ & $1.57 \cdot 10^{-8}$ \\
      0.16 &  $1.43 \cdot 10^{-7}$ &$3.79 \cdot 10^{-8}$ \\
      0.24 & $3.17 \cdot 10^{-7}$ &$7.72 \cdot 10^{-8}$ \\
      \hline
\end{tabular}
\caption{Conservation of energy (relative)}
\label{tab:landauB_energy}
\end{subtable}
\caption{Electrostatic test case: Conservation laws for the variational integrator with and without substeps. }
\end{table}

\subsubsection{An electromagnetic test case}

As a second example, we look at an electromagnetically dominated test case with initial distribution
\begin{align}
    &f(x, v_1,v_2)
    =\frac{1}{2\pi \sigma_1 \sigma_2} \exp \left(-\left( \frac{v_1^2}{2\sigma_1^2} + \frac{v_2^2}{2\sigma_2^2} \right) \right),
\end{align}
and initial magnetic field $B_0(x) = \beta_1 + \beta_2\cos(kx)$ on the domain $[0,\frac{2 \pi}{k})$. We choose the parameters to be $k = 1.25$, $\sigma_1 = \sqrt{2}\cdot 0.01$, $\sigma_2 = \sqrt{12}\sigma_1$, $\beta_1 = 20 \pi$, $\beta_2=0.001$. This test case is electromagnetic and a variation of the Weibel instability with a strong background field. The example is a variation of the test problem proposed in \cite{Bernier-et-al:2019}. We run a simulation until time 20 with 32 grid points, 100,000 particles, and spline basis functions of degree 3. The stability limit due to the splitting of the Maxwell's equation is at $\Delta t < (2 \pi)/(1.25 \times 32)\sqrt{17/42} \approx 0.09994$ in this case.

\begin{figure}
\centering
\begin{subfigure}{\figwidth\linewidth}
    \includegraphics[width=\figwidth\linewidth]{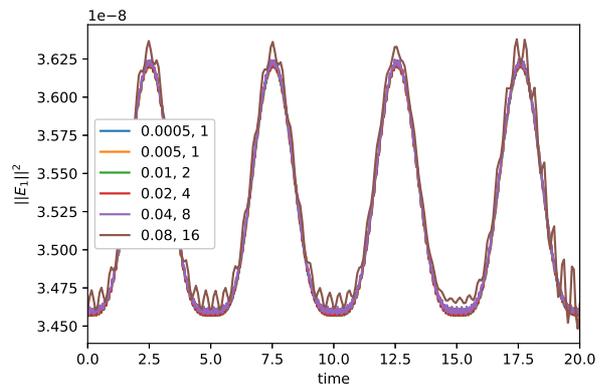}
    \caption{Variational subcycling.}
    \label{fig:weibelB_explicit_sub}
\end{subfigure}
~
\begin{subfigure}{\figwidth\linewidth}
    \includegraphics[width=\figwidth\linewidth]{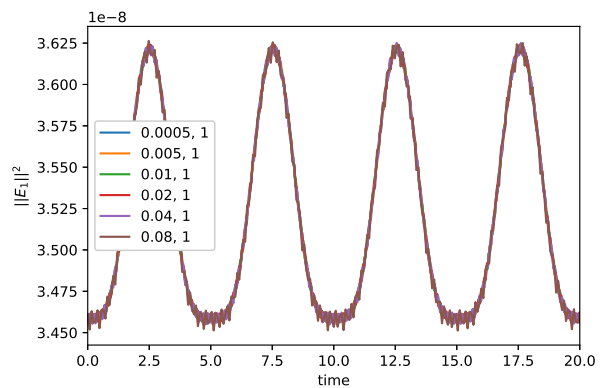}
    \caption{Variational algorithm.}
    \label{fig:weibelB_explicit}
\end{subfigure}
\caption{Electromagnetically dominated test case: Time evolution of $\|E_1\|^2$ for varying global time steps (given in the legends together with the number of substeps) with $B(x,0) = 2\pi 10$.}
\end{figure}

As in the previous example, we look at the oscillations in the first component of the electric energy. Again the cyclotron period is 0.1. 
Figure~\ref{fig:weibelB_explicit_sub} shows the results for various global time steps and fixed $\Delta \tau = 0.005$ and Figure~\ref{fig:weibelB_explicit} the results for the same global time steps but without subcycling. We can see that the qualitative behavior is the same as for the electrostatically dominated test case. Also Gauss' law is conserved to machine precision as shown in Table~\ref{tab:weibelB_Gauss}. The energy-conservation properties of the algorithm are summarized in Table~\ref{tab:weibelB_energy}. The fact that the energy conservation is better compared to the other test case, and that there is almost no difference comparing the scheme with and without subcycling, reflects the fact that the influence of the micro-scale is smaller in this test case with a smaller electric energy.
The number of Newton iterations needed (including the initial step) in this test case is only 1.7 on average with subcycling and increases up to 2.0 without subcycling for the largest time step.

\begin{table}
\begin{subtable}{0.48\textwidth}
\begin{tabular}{|c|c|c|}
\hline
     $\Delta t$  & with substeps & without substeps \\
     \hline 
      0.005 &                        &   $2.61 \cdot 10^{-15}$  \\
      0.01  & $2.65 \cdot 10^{-15}$  &   $2.26 \cdot 10^{-15}$  \\
      0.02  & $2.42 \cdot 10^{-15}$  &   $2.13 \cdot 10^{-15}$ \\
      0.04  & $2.10 \cdot 10^{-15}$  &    $2.31 \cdot 10^{-15}$ \\
      0.08  & $2.08 \cdot 10^{-15}$  &    $1.99 \cdot 10^{-15}$\\
      \hline
\end{tabular}
\caption{Conservation of Gauss' law}
\label{tab:weibelB_Gauss}
\end{subtable}
~
\begin{subtable}{0.48\textwidth}
\begin{tabular}{|c|c|c|}
\hline
     $\Delta t$&  with substeps & without substeps \\
     \hline
      0.005 &    & $3.98 \cdot 10^{-13}$ \\
      0.01 &   $7.98 \cdot 10^{-13}$ & $7.96 \cdot 10^{-13}$ \\
      0.02 &   $1.61 \cdot 10^{-12}$& $1.60 \cdot 10^{-13}$ \\
      0.04 &    $3.27 \cdot 10^{-12}$ & $3.24 \cdot 10^{-12}$ \\
      0.08 &    $1.47 \cdot 10^{-11}$ & $6.69 \cdot 10^{-12}$ \\
      \hline
\end{tabular}
\caption{Conservation of energy (relative)}
\label{tab:weibelB_energy}
\end{subtable}
\caption{Electromagnetic test case: Conservation laws for the variational integrator with and without substeps.}
\end{table}

\subsection{Enforced orbit averaging of electric impulse}
To investigate whether the root cause for the possible numerical oscillations is indeed in the way the electric field is evaluated in particle orbits, we will now enforce orbit averaging also for the electric-field contribution. 
For the indices $n$, we will use the following modified particle push 
\begin{align}\label{eq:discrete-el-x-integer_multistep-orbit-averaged}
    &m\frac{\bm{x}_{p,n+1/\mathcal{V}}-2\bm{x}_{p,n}+\bm{x}_{p,n-1/\mathcal{V}}}{(\Delta t/\mathcal{V})^ 2}
    \nonumber
    \\
    &=q\frac{\bm{x}_{n+1/\mathcal{V}}-\bm{x}_{p,n}}{\Delta t/\mathcal{V}}\nonumber\\
    &\qquad\qquad\times \int_0^1(1-\tau)(b_{n+1}^k+b_{\text{ext}}^k)\bm{W}^2_k(\bm{x}_{p}^{n,1}(\tau))d\tau
    \nonumber
    \\
    &+q\frac{\bm{x}_{p,n}-\bm{x}_{p,n-1/\mathcal{V}}}{\Delta t/\mathcal{V}}\nonumber
    \\
    &\qquad\qquad\times \int_0^1\tau (b_n^k+b_{\text{ext}}^k)\bm{W}^2_k(\bm{x}_{p}^{n, 0}(\tau))d\tau
    \nonumber
    \\
    &+q  e_n^j\bm{W}^1_j(\bm{x}_{p,n}),
\end{align}
and similarly for the indices $\nu$
\begin{align}\label{eq:discrete-el-x-half-integer_multistep-orbit-averaged}
    &m\frac{\bm{x}_{p,n+(\nu+1)/\mathcal{V}}-2\bm{x}_{p,n+\nu/\mathcal{V}}+\bm{x}_{p,n+(\nu-1)/\mathcal{V}}}{(\Delta t/\mathcal{V})^2}
    \nonumber
    \\
    &=q\frac{\bm{x}_{n+(\nu+1)/\mathcal{V}}-\bm{x}_{p,n+\nu/\mathcal{V}}}{\Delta t/\mathcal{V}}\nonumber
    \\
    &\qquad\qquad\times \int_0^1(1-\tau)(b_{n+1}^k+b_{\text{ext}}^k)\bm{W}^2_k(\bm{x}_{p}^{n,\nu+1}(\tau))d\tau
    \nonumber
    \\
    &+q\frac{\bm{x}_{p,n+\nu/\mathcal{V}}-\bm{x}_{p,n+(\nu-1)/\mathcal{V}}}{\Delta t/\mathcal{V}}\nonumber
    \\
    &\qquad\qquad\times \int_0^1\tau (b_{n+1}^k+b_{\text{ext}}^k)\bm{W}^2_k(\bm{x}_{p}^{n,\nu}(\tau))d\tau \nonumber
    \\
    &+q  e_n^j\bm{W}^1_j(\bm{x}_{p,n+\nu/\mathcal{V}}).
\end{align}
We stress that this particle push is not derived from an action principle and is not expected to provide bounded long-time energy behavior like the variational schemes, nor to conserve the multisymplectic two-form. For the field equations, we use the Amp\`ere and Gauss' law as described previously for they satisfy a charge conservation law regardless of how the particle orbits are sampled. 
\begin{figure}
\centering
    \includegraphics[width=\figwidth\linewidth]{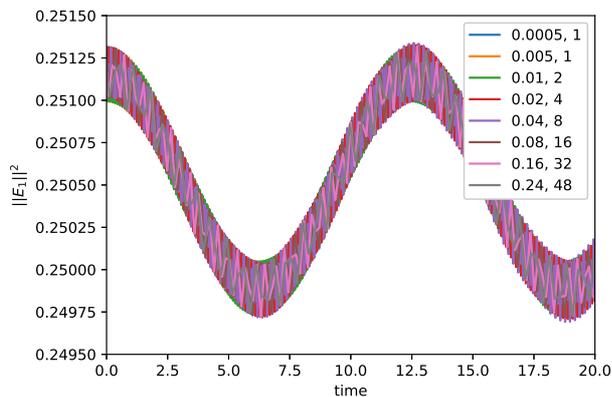}
\caption{Electrostatically dominated test case: Time evolution of $\|E_1\|^2$ for the subcycling algorithm with enforced electric-field orbit-averaging 
for various time steps (given in the legends together with the number of substeps).}\label{fig:landauB_eall}
\end{figure}

We then repeated the numerical tests from the previous section for the larger time steps using the above particle equations. For the electrostatically dominated test case, Figure~\ref{fig:landauB_eall} shows the evolution of the first component of the electric energy as a function of time with the new algorithm. We observe that the algorithm samples the energy curve quite accurately even if the global step size is increased beyond the cyclotron period. This indicates that finding a variational algorithm that properly orbit-averages the electric field impact should work well.

%% file: explicit_push.tex
\section{Explicit field solve with an explicit subcycling of the particle push}\label{sec:explicit-push}
So far we have discussed the introduction of subsycling to the pioneering algorithm~\cite{Squire-Qin-Tang-PIC:2012PhPl}. The more recent GEMPIC algorithms, however, largely are based on a particularly clever choice of representing the interaction term in the discrete action: instead of using a straight line in cartesian space, a zigzagging path and rectilinear meshes allow explicit particle push~\cite{Xiao-et-al:2018PlST}. Exploiting the zigzagging path, also the subcycling push can be made explicit. 

To prevent an excessive use of symbols, here we focus on the particle-relevant part of the action suppressing some indices to demonstrate that explicit subcycling is possible. Essential role is played by the following maps
\begin{align}
    \begin{bmatrix}
    \bm{x}_{\text{zig}}(\bm{x}_1,\bm{x}_2,\tau)\\
    \bm{y}_{\text{zig}}(\bm{x}_1,\bm{x}_2,\tau)\\
    \bm{z}_{\text{zig}}(\bm{x}_1,\bm{x}_2,\tau)
    \end{bmatrix}
    &=
    \begin{bmatrix}
    (x_1+\tau(x_2-x_1),y_1,z_1)\\
    (x_2,y_1+\tau(y_2-y_1),z_1)\\
    (x_2,y_2,z_1+\tau(z_2-z_1)),
    \end{bmatrix},
\end{align}
and a rectilinear mesh is required so that the vector valued basis functions can be represented component-wise along each coordinate direction. An interested reader may consult for example the Appendix A in Ref.~\cite{Xiao-et-al:2018PlST}. Effectively one needs a representation of the basis where, e.g., the discrete vector potential in cartesian coordinates becomes $\bm{A}=a^{j,x}\bm{W}_{j,x}^1+a^{j,y}\bm{W}_{j,y}^1+a^{j,z}\bm{W}_{j,z}^1$ and can be split along the coordinate axes to the related components $(A_x,A_y,A_z)$. The code within the \verb|SeLaLib| \cite{selalib} package with spline basis is implemented exactly in this manner.

Once the prerequisites are met, the subcycling can then be introduced by discretizing the single-particle relevant part of the action according to
\begin{align}
    &S_{n,n+1}[\bm{x}_n, \bm{x}_{n+1/\mathcal{V}},\ldots ,\bm{x}_{n+1},\bm{A}_{n+1},\phi_n]\nonumber\\
    &=\sum_{\nu=1}^{\mathcal{V}}\Biggr[q(x_{n+\nu/\mathcal{V}}-x_{n+(\nu-1)/\mathcal{V}})\nonumber
    \\
    &\qquad\int_0^1A_{x,n+1}(\bm{x}_{\text{zig}}(\bm{x}_{n+(\nu-1)/\mathcal{V}},\bm{x}_{n+\nu/\mathcal{V}},\tau))d\tau\nonumber
    \\
    &+q(y_{n+\nu/\mathcal{V}}-y_{n+(\nu-1)/\mathcal{V}})\nonumber
    \\
    &\qquad \int_0^1A_{y,n+1}(\bm{y}_{\text{zig}}(\bm{x}_{n+(\nu-1)/\mathcal{V}},\bm{x}_{n+\nu/\mathcal{V}},\tau))d\tau\nonumber
    \\
    &+q(z_{n+\nu/\mathcal{V}}-z_{n+(\nu-1)/\mathcal{V}})\nonumber
    \\
    &\qquad\int_0^1A_{z,n+1}(\bm{z}_{\text{zig}}(\bm{x}_{n+(\nu-1)/\mathcal{V}},\bm{x}_{n+\nu/\mathcal{V}},\tau))d\tau\Biggr]\nonumber
    \\
    &-q\phi_{n}(\bm{x}_n)\Delta t\nonumber
    \\
    &+\sum_{\nu=1}^{\mathcal{V}}\Biggr[ \frac{m}{2}\frac{(x_{n+\nu/\mathcal{V}}-x_{n+(\nu - 1)/\mathcal{V}})^2}{\Delta t/\mathcal{V}}
    \nonumber
    \\
    &+\frac{m}{2}\frac{(y_{n+\nu/\mathcal{V}}-y_{n+(\nu - 1)/\mathcal{V}})^2}{\Delta t/\mathcal{V}}\nonumber
    \\
    &+\frac{m}{2}\frac{(z_{n+\nu/\mathcal{V}}-z_{n+(\nu - 1)/\mathcal{V}})^2}{\Delta t/\mathcal{V}}\Biggr].
\end{align}

To derive the discrete equations of motion one proceeds by perturbing $x_{n+\nu/\mathcal{V}}\rightarrow x_{n+\nu/\mathcal{V}}+\delta x$ to get
\begin{align}
    &\frac{m}{q}\frac{x_{n+(\nu+1)/\mathcal{V}}-2x_{n+\nu/\mathcal{V}}+x_{n+(\nu-1)/\mathcal{V}}}{\Delta t/\mathcal{V}}\nonumber\\
    &=(y_{n+\nu/\mathcal{V}}-y_{n+(\nu-1)/\mathcal{V}})\nonumber
    \\
    &\qquad \int_0^1B_{z,n+1}(\bm{y}_{\text{zig}}(\bm{x}_{n+(\nu-1)/\mathcal{V}},\bm{x}_{n+\nu/\mathcal{V}},\tau))d\tau\nonumber\\
    &-(z_{n+\nu/\mathcal{V}}-z_{n+(\nu-1)/\mathcal{V}})\nonumber\\
    &\qquad \int_0^1B_{y,n+1}(\bm{z}_{\text{zig}}(\bm{x}_{n+(\nu-1)/\mathcal{V}},\bm{x}_{n+\nu/\mathcal{V}},\tau))d\tau,
\end{align}
perturbing $y_{n+\nu/\mathcal{V}}\rightarrow y_{n+\nu/\mathcal{V}}+\delta y$, one obtains
\begin{align}
    &\frac{m}{q}\frac{y_{n+(\nu+1)/\mathcal{V}}-2y_{n+\nu/\mathcal{V}}+y_{n+(\nu-1)/\mathcal{V}}}{\Delta t/\mathcal{V}}\nonumber\\
    &=-(x_{n+(\nu+1)/\mathcal{V}}-x_{n+\nu/\mathcal{V}})\nonumber\\
    &\qquad \int_0^1 B_{z,n+1}(\bm{x}_{\text{zig}}(\bm{x}_{n+\nu/\mathcal{V}},\bm{x}_{n+(\nu+1)/\mathcal{V}},\tau))d\tau\nonumber
    \\
    &+(z_{n+\nu/\mathcal{V}}-z_{n+(\nu-1)/\mathcal{V}})\nonumber\\
    &\qquad \int_0^1 B_{x,n+1}(\bm{z}_{\text{zig}}(\bm{x}_{n+(\nu-1)/\mathcal{V}},\bm{x}_{n+\nu/\mathcal{V}},\tau))d\tau,
\end{align}
and perturbing  $z_{n+\nu/\mathcal{V}}\rightarrow z_{n+\nu/\mathcal{V}}+\delta z$ provides
\begin{align}
    &\frac{m}{q}\frac{z_{n+(\nu+1)/\mathcal{V}}-2z_{n+\nu/\mathcal{V}}+z_{n+(\nu-1)/\mathcal{V}}}{\Delta t/\mathcal{V}}\nonumber\\
    &=(x_{n+(\nu+1)/\mathcal{V}}-x_{n+\nu/\mathcal{V}})\nonumber\\
    &\qquad \int_0^1B_{y,n+1}(\bm{x}_{\text{zig}}(\bm{x}_{n+\nu/\mathcal{V}},\bm{x}_{n+(\nu+1)/\mathcal{V}},\tau))d\tau\nonumber
    \\
    &-(y_{n+(\nu+1)/\mathcal{V}}-y_{n+\nu/\mathcal{V}})\nonumber\\
    &\qquad \int_0^1B_{x,n+1}(\bm{y}_{\text{zig}}(\bm{x}_{n+\nu/\mathcal{V}},\bm{x}_{n+(\nu+1)/\mathcal{V}},\tau))d\tau.
\end{align}

For the synchronizing steps, one perturbs $x_n\rightarrow x_n+\delta x$ which leads to
\begin{align}
    &\frac{m}{q}\frac{x_{n+1/\mathcal{V}}-2x_n+x_{n-1/\mathcal{V}}}{\Delta t/\mathcal{V}}\nonumber\\
    &=(y_{n}-y_{n-1/\mathcal{V}})\nonumber\\
    &\qquad \int_0^1B_{z,n}(\bm{y}_{\text{zig}}(\bm{x}_{n-1/\mathcal{V}},\bm{x}_{n},\tau))d\tau\nonumber
    \\
    &-(z_{n}-z_{n-1/\mathcal{V}})\nonumber\\
    &\qquad \int_0^1B_{y,n}(\bm{z}_{\text{zig}}(\bm{x}_{n-1/\mathcal{V}},\bm{x}_{n},\tau))d\tau\nonumber
    \\
    &+E_{x,n}(\bm{x}_n)\Delta t,
\end{align}
and $y_n\rightarrow y_n+\delta y$ which provides 
\begin{align}
    &\frac{m}{q}\frac{y_{n+1/\mathcal{V}}-2y_n+y_{n-1/\mathcal{V}}}{\Delta t/\mathcal{V}}\nonumber\\
    &=-(x_{n+1/\mathcal{V}}-x_{n})\nonumber\\
    &\qquad \int_0^1B_{z,n+1}(\bm{x}_{\text{zig}}(\bm{x}_{n},\bm{x}_{n+1/\mathcal{V}},\tau))d\tau\nonumber
    \\
    &+(z_{n}-z_{n-1/\mathcal{V}})\nonumber\\
    &\qquad \int_0^1B_{x,n}(\bm{z}_{\text{zig}}(\bm{x}_{n-1/\mathcal{V}},\bm{x}_{n},\tau))d\tau\nonumber
    \\
    &+E_{y,n}(\bm{x}_n)\Delta t,
\end{align}
while perturbing $z_n\rightarrow z_n+\delta z$ gives 
\begin{align}
    &\frac{m}{q}\frac{z_{n+1/\mathcal{V}}-2z_n+z_{n-1/\mathcal{V}}}{\Delta t/\mathcal{V}}\nonumber\\
    &=(x_{n+1/\mathcal{V}}-x_{n})\nonumber\\
    &\qquad \int_0^1B_{y,n+1}(\bm{x}_{\text{zig}}(\bm{x}_{n},\bm{x}_{n+1/\mathcal{V}},\tau))d\tau\nonumber
    \\
    &-(y_{n+1\mathcal{V}}-y_{n})\nonumber\\
    &\qquad \int_0^1B_{x,n+1}(\bm{y}_{\text{zig}}(\bm{x}_{n},\bm{x}_{n+1/\mathcal{V}},\tau))d\tau\nonumber
    \\
    &+E_{z,n}(\bm{x}_n)\Delta t.
\end{align}
These maps provide explicit update rules for particles and display the same behavior as the explicit subcycling scheme in the sense that the effect of the electric field is evaluated only once per subcycling period, i.e., during the synchronizing steps. The field solves remains the same as in the earlier algorithm, apart from remembering to perform the current deposits coordinate-direction wise via the $(\bm{x}_\text{zig},\bm{y}_\text{zig},\bm{z}_\text{zig})$ maps.

\begin{figure}
\centering
\begin{subfigure}{\figwidth\linewidth}
    \includegraphics[width=\figwidth\linewidth]{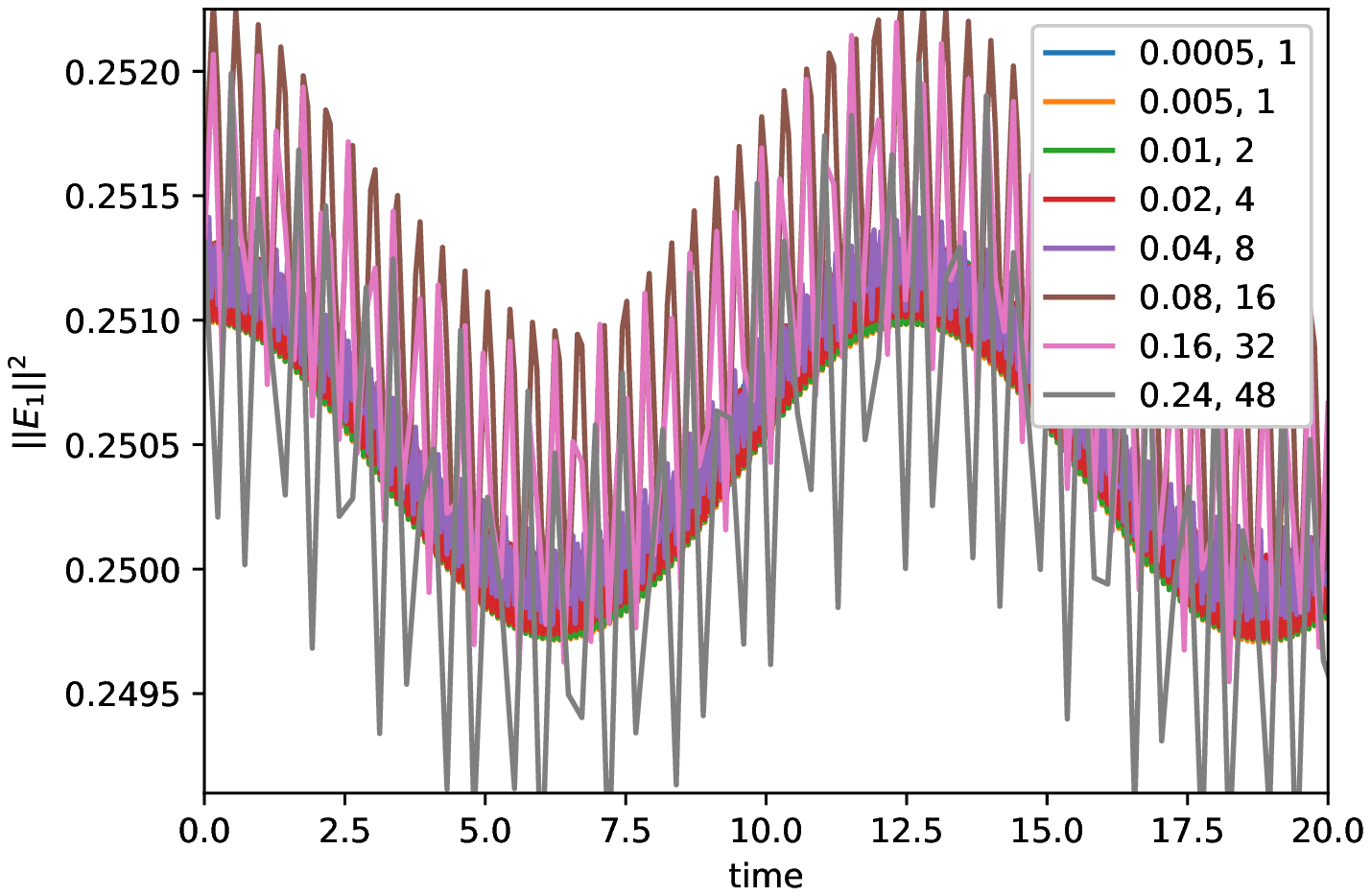}
    \caption{zigzag with subcycling ($\Delta \tau = 0.005$).}
    \label{fig:zig_sub20_landau}
\end{subfigure}
~
\begin{subfigure}{\figwidth\linewidth}
    \includegraphics[width=\figwidth\linewidth]{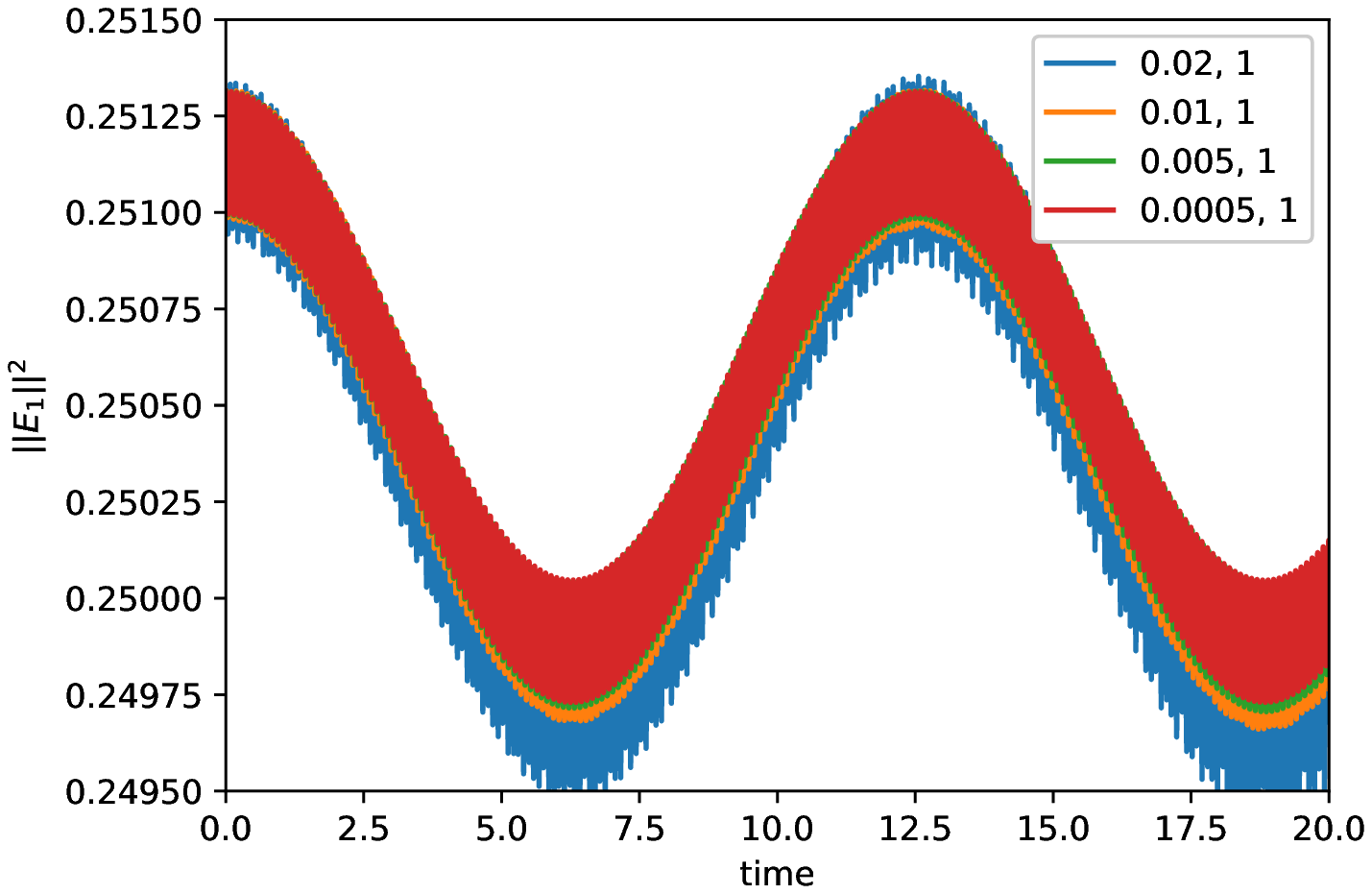}
    \caption{zigzag no subcycling.}
    \label{fig:zig_landau}
\end{subfigure}
\caption{Electrostatically dominated test case: Time evolution of $\|E_1\|^2$ for varying global time steps (given in the legends together with the number of substeps) with $B(x,0) = 2\pi 10$.}\label{fig:zigzag_landau}
\end{figure}

\begin{figure}
\centering
\includegraphics[width=\figwidth\linewidth]{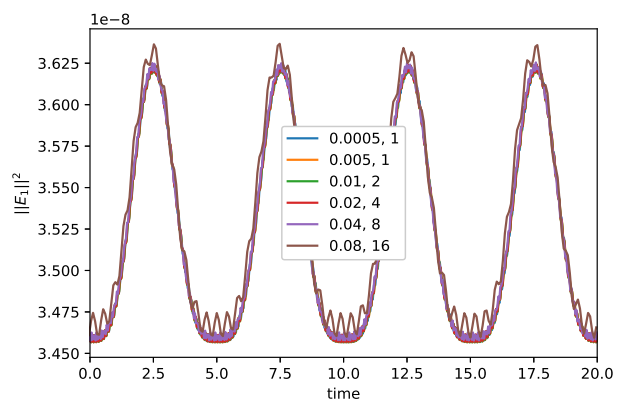}
\caption{Electromagnetically dominated test case, zigzag scheme with subcycling ($\Delta \tau = 0.005$): Time evolution of $\|E_1\|^2$ for varying global time steps (given in the legends together with the number of substeps) with $B(x,0) = 2\pi 10$.}
\label{fig:weibelB_zig}
\end{figure}

Let us revisit the electrostatically dominated test case but use the exlicit particle subcycling instead. 
Figure \ref{fig:zig_sub20_landau} shows the results with the same parameters as in Figure \ref{fig:explicit_sub20_landau}. We observe that the results are very similar, i.e., also with the explicit particle push very accurate results are obtained until $\Delta t = 0.04$ and the macroscopic behavior is approximately covered until the stability limit of the Maxwell solver. Moreover, the cyclotron oscillations are even covered a bit more accurately in this example by the zigzagging scheme as a zoom into the curves would reveal. On the other hand, if we do not use any substeps, the algorithm becomes unstable already for $\Delta t = 0.04$---Figure \ref{fig:zig_landau} displays only the stable step sizes. Moreover, the solution already with $\Delta t = 0.02$ covers the macroscopic properties less accurately than when subcycling is used.  Figure \ref{fig:weibelB_zig} shows the results for the electromagnetically dominated test case again with $\Delta \tau = 0.005$, i.e., 20 substeps per cyclotron period. The results are again very good except for the largest global time step of $\Delta t = 0.08$. Also, in this case, a slight deviation of the result is visible for $\Delta t = 0.04$.

We conclude that the explicit substepping with the zigzag algorithm not only allows for an accurate resolution of the phase with a moderate number of substeps but also comes with an increased stability domain in terms of the global time step in comparison to using the zigzag algorithm with no substeps. Tables \ref{tab:zigzag} and \ref{tab:zigzag_weibelB} show the conservation properties of the algorithm for the two test cases.

\begin{table}
\begin{subtable}{0.48\textwidth}
\begin{tabular}{|c|c|c|}
\hline
     $\Delta t$  & with substeps & without substeps \\
     \hline 
      0.005 &   $1.27 \cdot 10^{-14}$ & $5.29 \cdot 10^{-14}$ \\
      0.01 &    $1.28 \cdot 10^{-14}$ & $6.52 \cdot 10^{-14}$ \\
      0.02 &    $1.23 \cdot 10^{-14}$& $3.92 \cdot 10^{-14}$ \\
      0.04 &    $1.12 \cdot 10^{-14}$ & --- \\
      0.08 &    $1.25 \cdot 10^{-14}$ & --- \\
      0.16 &    $1.22 \cdot 10^{-14}$ & --- \\
      0.24 &    $1.05 \cdot 10^{-14}$ & --- \\
      \hline
\end{tabular}
\caption{Conservation of Gauss' law}
\label{tab:landauB_Gauss_zigzag}
\end{subtable}
~
\begin{subtable}{0.48\textwidth}
\begin{tabular}{|c|c|c|}
\hline
     $\Delta t$&  with substeps & without substeps \\
     \hline
      0.005 &  $2.56 \cdot 10^{-5}$ & $2.56 \cdot 10^{-5}$ \\
      0.01 &   $2.56 \cdot 10^{-5}$ & $1.11 \cdot 10^{-4}$ \\
      0.02 &   $2.56 \cdot 10^{-5}$ & $6.60 \cdot 10^{-4}$ \\
      0.04 &   $2.56 \cdot 10^{-5}$ & --- \\
      0.08 &   $2.56 \cdot 10^{-5}$ & --- \\
      0.16 &   $2.56 \cdot 10^{-5}$ & --- \\
      0.24 &   $2.57 \cdot 10^{-5}$ & ----\\
      \hline
\end{tabular}
\caption{Conservation of energy (relative)}
\label{tab:landauB_energy_zigzag}
\end{subtable}
\caption{Electrostatic test case: Conservation laws for the variational integrator with explicit zigzagging trajectories with and without substeps.}\label{tab:zigzag}
\end{table}

\begin{table}
\begin{subtable}{0.48\textwidth}
\begin{tabular}{|c|c|c|}
\hline
     $\Delta t$  & with substeps & without substeps \\
     \hline 
      0.005 &                       &    $2.88 \cdot 10^{-15}$  \\
      0.01  & $2.68 \cdot 10^{-15}$  &   $2.98 \cdot 10^{-15}$  \\
      0.02  & $2.46 \cdot 10^{-15}$  &   $2.63 \cdot 10^{-15}$ \\
      0.04  & $2.42 \cdot 10^{-15}$ &    --- \\
      0.08  & $1.97 \cdot 10^{-15}$ &    ---\\
      \hline
\end{tabular}
\caption{Conservation of Gauss' law}
\label{tab:weibelB_Gauss_zig}
\end{subtable}
~
\begin{subtable}{0.48\textwidth}
\begin{tabular}{|c|c|c|}
\hline
     $\Delta t$&  with substeps & without substeps \\
     \hline
      0.005 &                         & $2.87 \cdot 10^{-13}$ \\
      0.01 &    $2.87 \cdot 10^{-7}$ & $3.15 \cdot 10^{-13}$ \\
      0.02 &    $2.87 \cdot 10^{-7}$ & $4.94 \cdot 10^{-13}$ \\
      0.04 &    $2.87 \cdot 10^{-7}$ & --- \\
      0.08 &    $2.87 \cdot 10^{-7}$ & --- \\
      \hline
\end{tabular}
\caption{Conservation of energy (relative)}
\label{tab:weibelB_energy_zig}
\end{subtable}
\caption{Electromagnetic test case: Conservation laws for the variational integrator with explicit ziggagging trajectories with and without substeps.}\label{tab:zigzag_weibelB}
\end{table}

%% file: implicit-scheme.tex
\section{An implicit scheme with full orbit-averaging and electromagnetic gauge invariance}\label{sec:implicit}
To find a variational scheme that would succeed in fully orbit-averaging the particle trajectories, we suggest a temporal discretization that appears to lead to an implicit scheme. Essentially, we have learned that the key likely is in handling the interaction term in the action integral in a manner that properly allows one to perform integration by parts in time, instead of the summation-by-parts trick that works nicely without subcycling and apparently with certain limitations together with subcycling as described in Secs.~\ref{sec:explicit} and~\ref{sec:explicit-push}. It remains to be seen whether an explicit field-solve strategy, succeeding in both proper orbit-averaging and electromagnetic gauge-invariance, is possible. We also introduce arbitrary time steps for it might be useful in near future for adaptive temporal integration. 

\subsection{Polyline particle trajectories}
We will now assume that the particle trajectories form a polyline between different time instances and shall respect this assumption in the discretization. We partition the interval $[t_{\text{i}},t_{\text{f}}]$ into multiple arbitrary intervals $t_{\text{i}}=t_0<t_1<...<t_{m-1}<t_m<t_{m+1}<...<t_{\text{f}}$ and during each interval, the particle trajectory is expressed as
\begin{align}\label{eq:implicit-polyline}
    \bm{x}_p^{m,m+1}(t)&=\bm{x}_{p,m}+\frac{t-t_m}{t_{m+1}-t_m}(\bm{x}_{p,m+1}-\bm{x}_{p,m}),\\
    \dot{\bm{x}}_p^{m,m+1}(t)&=\frac{\bm{x}_{p,m+1}-\bm{x}_{p,m}}{t_{m+1}-t_m},
\end{align}
making sure that the time derivative is consistent with the trajectory. 
Substituting these expressions into the action, we find the following particle-relevant contribution over the interval $[t_m,t_{m+1}]$ 
\begin{align}
    &S^p_{m,m+1}[\mathbbm{a}(t),\phi(t),\mathbbm{x}_m,\mathbbm{x}_{m+1}]\nonumber
    \\
    &=q\int_{t_m}^{t_{m+1}} (a^j(t)+a_{\text{ext}}^j)\bm{W}^1_j(\bm{x}_p^{m,m+1}(t))\cdot\dot{\bm{x}}_p^{m,m+1}(t)dt\nonumber
    \\
    &-q\int_{t_m}^{t_{m+1}}\phi^i(t)W_i^0(\bm{x}_p^{m,m+1}(t))dt\nonumber
    \\
    &+\frac{1}{2}m\int_{t_m}^{t_{m+1}}|\dot{\bm{x}}_p^{m,m+1}|^2 dt.
\end{align}
Perturbing the particle polylines into $\mathbbm{x}_m+\epsilon\delta\mathbbm{x}_m$ and minimizing the action with respect to the variations in the particle positions, we obtain the following discrete Euler--Lagrange condition for each particle
\begin{align}
    &\partial_{\epsilon}|_{\epsilon=0}S^p_{m,m+1}[\bm{x}_{p,m}+\epsilon\delta\bm{x}_{p,m}]
    \nonumber\\
    &+\partial_{\epsilon}|_{\epsilon=0}S^p_{m-1,m}[\bm{x}_{p,m}+\epsilon\delta\bm{x}_{p,m}]=0.
\end{align}
Written explicitly, this corresponds to
the following discrete Euler--Lagrange condition
\begin{align}\label{eq:implicit-particle-push}
    &m\frac{\bm{x}_{p,m+1}-\bm{x}_{p,m}}{t_{m+1}-t_m}-m\frac{\bm{x}_{p,m}-\bm{x}_{p,m-1}}{t_m-t_{m-1}}\nonumber
    \\
    &=q\frac{\bm{x}_{p,m+1}-\bm{x}_{p,m}}{t_{m+1}-t_m}\nonumber\\
    &\qquad\times\int_{t_m}^{t_{m+1}} \frac{t_{m+1}-t}{t_{m+1}-t_m}(b^k(t)+b_{\text{ext}}^k)\bm{W}^2_k(\bm{x}_p^{m,m+1}(t))dt
    \nonumber 
    \\
    &+q\frac{\bm{x}_{p,m}-\bm{x}_{p,m-1}}{t_m-t_{m-1}}\nonumber\\
    &\qquad\times\int_{t_{m-1}}^{t_{m}} \frac{t-t_{m-1}}{t_m-t_{m-1}}(b^k(t)+b_{\text{ext}}^k)\bm{W}^2_k(\bm{x}_p^{m-1,m}(t))dt
    \nonumber 
    \\
    &+q\int_{t_{m-1}}^{t_{m}}\frac{t-t_{m-1}}{t_m-t_{m-1}} e^j(t)\bm{W}^1_j(\bm{x}_p^{m-1,m}(t))dt
    \nonumber 
    \\
    &+q\int_{t_m}^{t_{m+1}}\frac{t_{m+1}-t}{t_{m+1}-t_m}e^j(t)\bm{W}^1_j(\bm{x}_p^{m,m+1}(t))dt,
\end{align}
where we have associated $e^j(t)=-\dot{a}^j(t)-\phi^i(t)\grad_i^j$.

In deriving the expression \eqref{eq:implicit-particle-push}, it was necessary to request time-continuity for $a^j(t)$ but not for $\phi^i(t)$. Hence we can imagine a piece-wise time-constant electric field $e^j(t)=-\dot{a}^j(t)-\grad_i^j\phi^i(t)$. The magnetic field $b^k(t)$ appearing in the particle equation, however, has to be at least piece-wise linear in time for it needs to be compatible with the requirement of at least piecewise linear $a^j(t)$ in the interaction part of the action.

\subsection{Polyline \texorpdfstring{$a(t)$}{TEXT} and piece-wise constant \texorpdfstring{$\phi(t)$}{TEXT}}

Next we partition the interval $[t_{\text{i}},t_{\text{f}}]$ according to $t_{\text{i}}=t_0<t_1<...<t_{n-1}<t_n<t_{n+1}<...<t_{\text{f}}$, again with arbitrary intervals. During each interval $[t_{n},t_{n+1}]$ we assume the following behavior for the electromagnetic degrees of freedom
\begin{align}
    a^j_{n,n+1}(t)&=a^j_{n}+\frac{t-t_n}{t_{n+1}-t_n}(a^j_{n+1}-a^j_{n}),\\
    \phi^i_{n,n+1}(t)&=\phi^i_{n},
\end{align}
which implies that we can define the electric and magnetic field during the interval directly via the relations 
\begin{align}
    \label{eq:interpolated-b}
    b^k_{n,n+1}(t)&=a^j_{n,n+1}(t)\curl_j^k\nonumber
    \\
    &\equiv b^k_{n}+\frac{t-t_n}{t_{n+1}-t_n}(b^k_{n+1}-b^k_{n}), \\
    \label{eq:interpolated-e}
    e^j_{n,n+1}(t)&=-\frac{a^j_{n+1}-a^j_n}{t_{n+1}-t_{n}}-\phi^i_{n}\grad_i^j\nonumber
    \\
    &\equiv e^j_{n}.
\end{align}
The above discretizations satisfy the requirements for the magnetic field and potential to be at least time-continuous and the electric field at least piece-wise constant, thus being compatible with \eqref{eq:implicit-particle-push}.
The discretization also implies a form for the discrete Faraday law
\begin{align}\label{eq:implicit-Faraday-law}
    \frac{b^k_{n+1}-b^k_n}{t_{n+1}-t_{n}}=-e^j_{n}\curl_j^k.
\end{align}
Substituting these expressions into the action, we find the following electromagnetic-relevant contribution over the interval $[t_n,t_{n+1}]$
\begin{align}\label{eq:implicit-action-EM}
    &S^{\text{EM}}_{n,n+1}[\mathbbm{a}_{n},\mathbbm{a}_{n+1},\phi_{n},\mathbbm{x}(t)]
    \nonumber
    \\
    &=\frac{\varepsilon_0}{2}e_{n}^{j_1}M^1_{j_1,j_2}e_{n}^{j_2}(t_{n+1}-t_n)\nonumber
    \\
    &-\frac{\mu_0^{-1}}{2}\int_{t_{n}}^{t_{n+1}}(b^{k_1}_{n,n+1}(t)+b_{\text{ext}}^{k_1})M^2_{k_1k_2}(b^{k_2}_{n,n+1}(t)+b_{\text{ext}}^{k_2})dt
    \nonumber
    \\
    &+\sum_p\int_{t_{n}}^{t_{n+1}} q(a^j_{n,n+1}(t)+a_{\text{ext}}^j)\bm{W}^1_j(\bm{x}_p(t))\cdot\dot{\bm{x}}_p(t)dt
    \nonumber
    \\
    &-\sum_p\int_{t_{n}}^{t_{n+1}}q\phi_{n}^iW_i^0(\bm{x}_p(t))dt .
\end{align}

Next we perturb the degrees of freedom for the vector potential into $\mathbbm{a}_{n}+\epsilon\delta\mathbbm{a}_n$ and minimize the action with respect to the variations $\delta\mathbbm{a}_n$. This provides the discrete Euler--Lagrange equation
\begin{align}
    \partial_{\epsilon}|_{\epsilon=0}S^{\text{EM}}_{n,n+1}[\mathbbm{a}_n+\epsilon\delta \mathbbm{a}_n]+\partial_{\epsilon}|_{\epsilon=0}S^{\text{EM}}_{n-1,n}[\mathbbm{a}_{n}+\epsilon\delta \mathbbm{a}_{n}]=0,
\end{align}
and explicitly it provides the following discrete Amp\`ere equation
\begin{align}\label{eq:implicit-ampere}
    &\varepsilon_0M^1_{j_1,j_2}(e_{n}^{j_2}-e_{n-1}^{j_2})+J_{j+}^{n,n+1}+J_{j-}^{n-1,n}
    \nonumber
    \\
    &=\mu_0^{-1}\curl_{j_1}^{k_1}M_{k_1k_2}^2\left(\frac{1}{6}b_{n+1}^{k_2}+\frac{1}{3}b_{n}^{k_2}+\frac{1}{2}b_{\text{ext}}^{k_2}\right)(t_{n+1}-t_n)
    \nonumber
    \\
    &+\mu_0^{-1}\curl_{j_1}^{k_1}M_{k_1k_2}^2\left(\frac{1}{3}b_{n}^{k_2}+\frac{1}{6}b_{n-1}^{k_2}+\frac{1}{2}b_{\text{ext}}^{k_2}\right)(t_{n}-t_{n-1}),
\end{align}
where the discrete current densities are defined via the relations
\begin{align}
    \label{eq:implicit-current-plus}
    J_{j+}^{n,n+1}&=q\sum_p\int_{t_n}^{t_{n+1}}\tfrac{t_{n+1}-t}{t_{n+1}-t_{n}}\bm{W}^1_{j_1}(\bm{x}_p(t))\cdot\dot{\bm{x}}_{p}(t)dt,\\
    \label{eq:implicit-current-minus}
    J_{j-}^{n,n+1}&=q\sum_p\int_{t_{n}}^{t_{n+1}}\tfrac{t-t_{n}}{t_{n+1}-t_{n}}\bm{W}^1_{j_1}(\bm{x}_p(t))\cdot\dot{\bm{x}}_{p}(t)dt.
\end{align}
Note that in deriving the expression \eqref{eq:implicit-ampere}, it is enough to require continuity from $\bm{x}_p(t)$, while the corresponding $\dot{\bm{x}}_p(t)$ can be piece-wise constant. Hence this Amp\`ere equation and the equation for the particle motion \eqref{eq:implicit-particle-push} are fully compatible with each other. One only has to account for the fact that the instances $t_n$ and $t_m$ do not necessarily coincide. 

Finally, perturbing the degrees for the scalar potential to $\phi_{n}+\epsilon\delta\phi_{n}$ and extremizing the action with respect to arbitrary variations $\delta\phi_{n}$ according to
\begin{align}
    \partial_{\epsilon}|_{\epsilon=0}S_{n,n+1}^{\text{EM}}[\phi_{n}+\epsilon\delta\phi_{n}]=0
\end{align}
provides the discrete Gauss' law 
\begin{align}\label{eq:implicit-Gauss-law}
    \varrho_i^{n,n+1}=-\varepsilon_0\grad_i^jM^1_{j,j_2}e_{n}^{j_2},
\end{align}
where the discrete charge density is defined as
\begin{align}\label{eq:implicit-charge}
    \varrho_i^{n,n+1}=q\sum_p\int_{t_{n}}^{t_{n+1}}W^0_i(\bm{x}_p(t))\frac{dt}{t_{n+1}-t_{n}}.
\end{align}

\subsection{Gauge invariance and the charge-conservation law}
To demonstrate that the Gauss' law \eqref{eq:implicit-Gauss-law} serves only as an initial condition and that it is enough to advance the electric field degrees of freedom via the discrete Amp\`ere equation \eqref{eq:implicit-ampere}, we start from the electromagnetic gauge invariance.

We define a gauge transformation with a function
\begin{align}
    \chi^i_{n,n+1}(t)=\chi^i_{n}+\frac{t-t_{n}}{t_{n+1}-t_n}(\chi^i_{n+1}-\chi^i_{n}),
\end{align}
and change the discrete vector and scalar potentials according to
\begin{align}
    &a^j_{n,n+1}(t)\rightarrow a^j_{n,n+1}(t)+\chi^i_{n,n+1}(t)\grad_i^j,\\
    &\phi^i_{n,n+1}(t)\rightarrow \phi^i_{n,n+1}(t)-\dot{\chi}^i_{n,n+1}(t).
\end{align}
The discrete electric and magnetic field are trivially unchanged under these substitutions and the relevant part of the action then satisfies
\begin{align}
    &\sum_{n=0}^{N-1}S^{\text{EM}}_{n,n+1}[a^j_{n}+\grad_i^j\chi^i_{n},a_{n+1}+\grad_i^j\chi^i_{n+1},\nonumber\\
    &\qquad\qquad\qquad\phi_{n}-(\chi_{n+1}^i-\chi_{n})/(t_{n+1}-t_{n})]
    \nonumber
    \\
    &=\sum_{n=0}^{N-1}S^{\text{EM}}_{n,n+1}[a^j_{n},a_{n+1},\phi_{n}]\nonumber\\
    &+q\sum_p[\chi^i_{N}W^0_i(\bm{x}_p(t_{f}))-\chi^i_{0}W^0_i(\bm{x}_p(t_i))].
\end{align}
Proceeding as previously, i.e., differentiating the above relation with respect to $\chi_{n}^i$ for arbitrary $n\in \{1,\ldots, N-1\}$, provides the discrete charge conservation law
\begin{align}\label{eq:implicit-charge-conservation}
    &\grad_i^j\left(J_{j+}^{n,n+1}+J_{j-}^{n-1,n}\right)-\left(\varrho_i^{n,n+1}-\varrho_i^{n-1,n}\right)=0,
\end{align}
where the current and charge densities are as defined in the equations \eqref{eq:implicit-current-plus}, \eqref{eq:implicit-current-minus}, and \eqref{eq:implicit-charge}.

Assuming the discrete Gauss' law \eqref{eq:implicit-Gauss-law} to hold for $n-1$, it is then a straightforward task to use the Amp\`ere equation \eqref{eq:implicit-ampere} together with the charge conservation law \eqref{eq:implicit-charge-conservation} to obtain
\begin{align}\label{eq:gauss_law_implicit}
    \varrho_i^{n,n+1}&=\varrho_i^{n-1,n}-\grad_i^j\left(J_{j+}^{n,n+1}+J_{j-}^{n-1,n}\right)\nonumber
    \\
    &=-\varepsilon_0\grad_i^jM^1_{j,j_2}e_{n}^{j_2}.
\end{align}
This means that, if the Gauss' law holds initially, it will be satisfied for all times when we solve the electric field from the Amp\`ere equation. This result is fully analogous with the one we obtained for the algorithm with an explicit field solve in Sec.~\ref{sec:explicit}.

\subsection{Solver strategy and equal-step sequencing}
Next we propose one possible strategy to implement the implicit scheme as described above, using equal step sizes for all but the first global step and fixed-point iteration for the nonlinear solves. 
Letting $\mathcal{V}$ denote the number of particle subcycling steps per one global time step $\Delta t$, we define for $n=1,m=1$ 
\begin{align}
    t_{1}&=t_0+\Delta t/\mathcal{V},
\end{align}
while for $n>1,m>1$ we define
\begin{align}
    t_{n+1}&=t_n+\Delta t,\\
    t_{m+1}&=t_m+\Delta t/\mathcal{V}.
\end{align}
Introducing the index $\nu$ as in the explicit section, one could then interpret the time instances $t_m$ to correspond to $t_{n+\nu/\mathcal{V}}=t_n+\nu/\mathcal{V}\Delta t$, with $n=\lfloor (m+(\mathcal{V}-1))/\mathcal{V}\rfloor,\nu=\bmod(m,\mathcal{V})$ for $m>0$. In explaining our sequencing strategy we shall hence refer with $\bm{x}_{n+\nu/\mathcal{V}}$ to particle location at $t_m=t_{n+\nu/\mathcal{V}}$. 

The solution strategy proceeds by first setting up the simulation:
\begin{enumerate}
    \item Given $\mathbbm{x}_0,\mathbbm{v}_0$ as samples from the initial distribution, compute $\mathbbm{x}_1=\mathbbm{x}_0+\Delta t/\mathcal{V}\mathbbm{v}_0$
    \item Given $\mathbbm{x}_0,\mathbbm{x}_1$, compute $\varrho^{0,1}_i$ from \eqref{eq:implicit-charge}, solve $\mathbbm{e}_0$ from \eqref{eq:implicit-Gauss-law}, and compute $J_{j-}^{0,1}$ from \eqref{eq:implicit-current-minus}.
    \item Given $\mathbbm{e}_0,\mathbbm{b}_0$, solve $\mathbbm{b}_1$ from \eqref{eq:implicit-Faraday-law}
\end{enumerate}
Next we assume to be in possession of $\mathbbm{e}_{n-1}$, $\mathbbm{b}_{n-1},\mathbbm{b}_n$, $J_{j-}^{n-1,n}$ and $\mathbbm{x}_{n-1/\mathcal{V}},\mathbbm{x}_{n}$, which is now obviously true for $n=1$. To advance the index $n$, we may proceed by following an iterative strategy:
\begin{enumerate}
    \item Guess $\mathbbm{e}_n$
    \item Given $\mathbbm{e}_n,\mathbbm{b}_n$, compute $\mathbbm{b}_{n+1}$ from \eqref{eq:implicit-Faraday-law}
    \item Given $\mathbbm{x}_{n-1/\mathcal{V}},\mathbbm{x}_n$ compute $\mathbbm{x}_{n+1/\mathcal{V}}$ using $\mathbbm{e}_{n-1},\mathbbm{e}_n$ and $\mathbbm{b}_{n-1},\mathbbm{b}_n,\mathbbm{b}_{n+1}$ from \eqref{eq:implicit-particle-push}.
    \item Given $\mathbbm{x}_n,\mathbbm{x}_{n+1/\mathcal{V}}$ compute $\mathbbm{x}_{n+2/\mathcal{V}},...,\mathbbm{x}_{n+1}$ using $\mathbbm{e}_n,\mathbbm{b}_n,\mathbbm{b}_{n+1}$ from \eqref{eq:implicit-particle-push}.
    \item Given $\mathbbm{x}_{n},...,\mathbbm{x}_{n+1}$ compute $J_{j+}^{n,n+1}$ from \eqref{eq:implicit-current-plus}
    \item Given $J_{j-}^{n-1,n},J_{j+}^{n,n+1}$ and $\mathbbm{e}_{n-1},\mathbbm{b}_{n-1},\mathbbm{b}_n,\mathbbm{b}_{n+1}$, solve $\mathbbm{e}_n$ from \eqref{eq:implicit-ampere}
    \item Iterate the steps 2--6 until $\mathbbm{e}_n$ converges.
\end{enumerate}
Note that in performing the particle push, the expressions $b^k(t),e^j(t)$ appearing in \eqref{eq:implicit-particle-push} are given by \eqref{eq:interpolated-b} and \eqref{eq:interpolated-e}, and similarly the particle trajectory appearing in the expressions for the current densities is the polyline~\eqref{eq:implicit-polyline}.




\subsection{Stability limit of the Maxwell solver}

Even though Maxwell's equations are solved implicitly in this scheme, the Maxwell part is not unconditionally stable. We can perform a stability analysis of the scheme in 1d2v phase-space following Appendix A2 of \cite{Kormann2019}. From this analysis, we get a stability condition of $\Delta t < \sqrt{\frac{17}{14}}\Delta x$ for the case of a cubic spline finite element solver (that we use in our experiments). This means the stability limit is relaxed by a factor $\sqrt{3}$ compared to the explicit scheme.

\subsection{Numerical tests}

We again repeat the previous experiments. The nonlinear iteration for each particle is stopped at a tolerance of $10^{-10}$ as before and the nonlinear iteration over the fields is considered converged at a tolerance of $10^{-13}$.

\begin{figure}
\centering
\begin{subfigure}{\figwidth\linewidth}
    \includegraphics[width=\figwidth\linewidth]{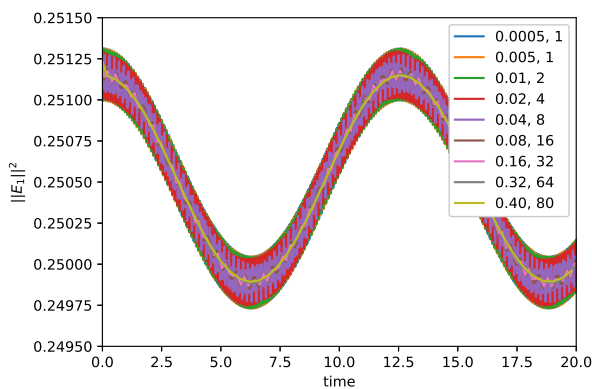}
    \caption{Electrostatically dominated.}
    \label{fig:landauB_implicit}
\end{subfigure}
~
\begin{subfigure}{\figwidth\linewidth}
    \includegraphics[width=\figwidth\linewidth]{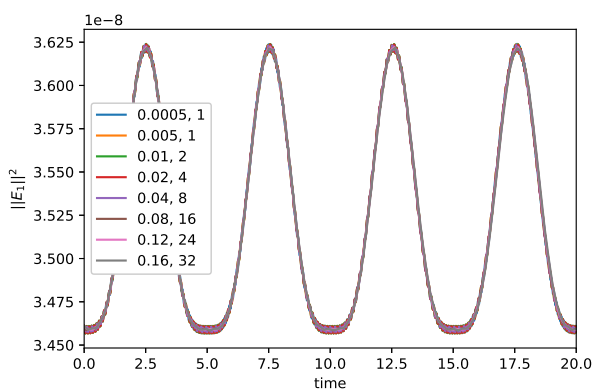}
    \caption{Electromagnetically dominated.}
    \label{fig:weibelB_implicit}
\end{subfigure}
\caption{Implicit subcycling scheme: Time evolution of $\|E_1\|^2$  with various time steps (given in the legends).}\label{fig:implicit}
\end{figure}

Figure~\ref{fig:implicit} shows the evolution of the first component of the electric energy for the two test cases studied previously, now obtained with the implicit scheme. We observe that the implicit scheme with subcycling provides a solution that accurately follows the macro-scale behavior of the solution for all of the time steps up to the stability limit of the scheme. Unlike the non-variational explicit enforced orbit-averaging scheme, the implicit scheme also appears to "step over" the fastest time scales and provides a rather smooth, averaged overall behavior when using time steps longer than the cyclotron period.

Again, we see from the second column of Tables~\ref{tab:landauB_implicit} and \ref{tab:weibelB_implicit} that the Gauss law is satisfied to machine precision. From the third column of Table~\ref{tab:landauB_implicit}, we see that the relative energy error is almost independent of the global time step for these simulations with constant $\Delta \tau$. 

Regarding computational performance, the number of Newton updates for solving the individual particle push is slightly increased compared to the scheme with the explicit field solve, to about 3.9 in the electrostatically dominated test case and 2.0 in the electromagnetically dominated test case. The number of global iterations in the electromagnetically dominated case, to perform the field solves, is 2.0 on average for all global time steps but $\Delta t = 0.16$, where it is 2.4. In the electrostatically dominated case, the corresponding number of global iterations is similar, namely 2.0 for $\Delta t = (0.005, 0.01, 0.02)$, 2.7 for $\Delta t = 0.08$, and 3.0 for $\Delta t = (0.16,0.32,0.4)$.

\begin{table}
\begin{subtable}{0.48\textwidth}
\begin{tabular}{|c|c|c|}
\hline
     $\Delta t$ & Gauss' law & Energy \\
     \hline
      0.005  &   $8.91 \cdot 10^{-15}$ & $1.48 \cdot 10^{-9}$ \\
      0.01  &  $1.26 \cdot 10^{-14}$ & $2.21 \cdot 10^{-9}$ \\
      0.02  &  $1.68 \cdot 10^{-14}$ & $3.82 \cdot 10^{-9}$ \\
      0.04 &  $2.26 \cdot 10^{-14}$ & $6.97 \cdot 10^{-9}$ \\
      0.08 &  $2.47 \cdot 10^{-14}$ & $8.40 \cdot 10^{-9}$ \\
      0.16 &  $3.54 \cdot 10^{-14}$ & $7.94 \cdot 10^{-9}$ \\
      0.32 &  $3.94 \cdot 10^{-14}$ & $7.81 \cdot 10^{-9}$ \\
      0.40 &  $5.48 \cdot 10^{-14}$ & $7.57 \cdot 10^{-9}$ \\
      \hline
\end{tabular}
\caption{Electrostatically dominated test case.}
\label{tab:landauB_implicit}
\end{subtable}
~
\begin{subtable}{0.48\textwidth}
\begin{tabular}{|c|c|c|}
\hline
     $\Delta t$ &   Gauss' law & Energy \\
     \hline
      0.005 & $4.32 \cdot 10^{-15}$ & $4.01 \cdot 10^{-13}$ \\
      0.01 & $4.97 \cdot 10^{-15}$ & $8.09 \cdot 10^{-13}$ \\
      0.02 & $7.44 \cdot 10^{-15}$ & $1.65 \cdot 10^{-12}$ \\
      0.04 & $8.53 \cdot 10^{-15}$ & $3.43 \cdot 10^{-12}$ \\
      0.08 & $1.10 \cdot 10^{-14}$ & $7.38 \cdot 10^{-12}$ \\
      0.12 & $8.59 \cdot 10^{-15}$ & $1.19 \cdot 10^{-11}$ \\
      0.16 & $1.14 \cdot 10^{-14}$ & $1.68 \cdot 10^{-11}$ \\
      \hline
\end{tabular}
\caption{Electromagnetically dominated test case.}
\label{tab:weibelB_implicit}
\end{subtable}
\caption{Implicit subcycling scheme: Conservation laws for different  time steps.}\label{tab:implicit}
\end{table}

%% file: hpc.tex
\section{Discussion on the computational efficiency and 
stability}\label{sec:hpc}
Finally, let us provide some insight into the computational efficiency and stability of the novel schemes, at least on a theoretical level. To properly access the performance computationally, a platform specific high-performance implementation would be necessary. Efficient and hardware-aware implementation of particle-in-cell codes, in particular containing integrals over intervals of varying length, is a research topic on its own and beyond the scope of this work. 

The main supposed computational benefit in using a subcycling scheme, in comparison to no subcycling at all, is the fact that the global step size for the field solves can be relaxed from the step size for electrons and that the different ion species can have their own time steps characterized by the ion cyclotron timescales---one needs to be mindful of the fundamental limits set by the CFL-condition and the plasma frequency, though. In order to get an idea on the gain of the subcycling, let us assume that the total computational cost is dominated by the particle push, as it often tends to be in a particle-in-cell implementation. 

If we can increase the step size for all $S$ ion species to $M$-times the time step of electrons, the computational complexity of the subcycling methods applying explicit field solves, compared to the same algorithms with no subcycling at all, behaves as
\begin{align}
    \frac{M+S}{(S+1)M}.
\end{align}
With only one species of ions ($S=1$) and, say, eight subcycling steps $(M=8)$ which still appeared to provide reasonable accuracy in our tests in Secs.~\ref{sec:explicit} and~\ref{sec:explicit-push}, the computational complexity would be reduced to 0.56 times the original, already close to the ideal 0.5. If the number of ion species is increased to $S=4$, the computational complexity is reduced to 0.3 times the original. This means that the subcycling schemes with explicit field solves have excellent potential to reduce the computational complexity in simulations of especially multiple species in magnetized plasmas. 
This observation is emphasized by the fact that the modifications introduced by the subcycling to the corresponding existing variational methods without subcycling are minimal.
    
Analysis of the fully implicit scheme against the explicit field solve schemes is not that much different: the implicit field solve requires an iterative approach which increases the number of necessary evaluations of the particle orbits and, if the comparison is done against the fully explicit scheme, also the number of iterations necessary for a single-particle push need to be accounted for. If we denote the total number of iterations by a factor of $I$ we have a new estimate for the relative complexity
\begin{align}
    \frac{I(M+S)}{(S+1)M}.
\end{align}
For example, in the electrostatically dominated simulation with the global step of $\Delta t=0.32$ and the number of subcycling steps of 64, we had to evaluate 3 fixed-point iterations on average and 4 iterations for the particle. If compared to the fully explicit scheme with no sybcycling, we then take $I=12$. Assuming only one ion species, the complexity of the implicit scheme would be 6.09, but with $S=4$, this would be reduced to 2.55. If compared to the explicit field solve method which uses the implicit particle push, one would use $I=3$ and relative complexities in the cases of one ion species and $S=4$ would be 1.52 and 0.64 respectively. Hence, based on these estimates, we could expect the new fully implicit scheme to reach a breakeven even against the fully explicit scheme when there are approximately 10 times more ion marker particles than electron markers.  

Finally, both the explicit and implicit sybcycling methods are expected to increase the arithmetic intensity as all of the substeps during one subcycling period can be performed without updating the field parameters. This should render the subcycling algorithms to behave favorably on modern computer architecture. Assessing this property thoroughly would, however, require high-performance implementations of the methods.

Regarding the performance of the algorithms in long-time simulations, variational methods (and Hamiltonian splitting schemes) in general are the best tools available. This is typically merited to the conservation of the multisymplectic two-form and the good behavior of energy in the sense that it is bounded with the bounds depending on the time step size. The analyses are typically performed for synchronous integrators, but examples exist also for asynchronous variational integrators \cite{Lew-et-al:2003ArRMA}. We anticipate that such rigorous analysis could be extended also to our subcycling schemes but this is left to a future study. 

\begin{figure}
\centering
\begin{subfigure}{\figwidth\linewidth}
    \includegraphics[width=\figwidth\linewidth]{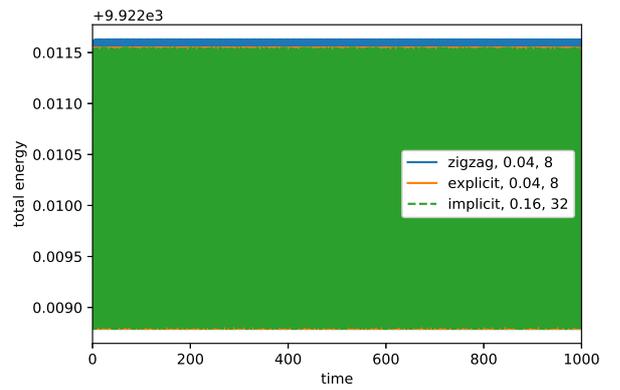}
    \caption{Variational schemes.}
    \label{fig:long_time_var}
\end{subfigure}
~
\begin{subfigure}{\figwidth\linewidth}
    \includegraphics[width=\figwidth\linewidth]{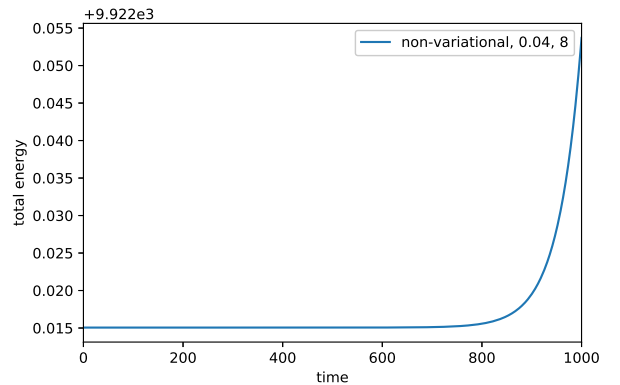}
    \caption{Non-variational scheme.}
    \label{fig:long_time_nonvar}
\end{subfigure}
\caption{Evolution of the total energy over 1000 time units with the explicit and implicit field solve, the zigzagging scheme (a), and the non-variational explicit orbit-averaging scheme (b). The number of $\Delta t, \mathcal{V}$ is given in the legend.}\label{fig:long_time}
\end{figure}

It would likely be possible to perform also the so-called backwards error analysis using the flow maps of the numerical schemes to find a Taylor series expression for the Hamiltonian the discrete flow map conserves exactly. This procedure has been left for future analysis, though, for we expect it to be somewhat more complicated a procedure than the backwards error analysis of synchronous integrators. Instead, we have simply run some of the simulations over prolonged intervals, corresponding to 1000.0 time units. Figure \ref{fig:long_time_var} shows the time evolution of the total energy for the electromagnetically dominated test case using a coarse resolution of 16 grid points and 50,000 particles. We observe the typical behavior for variational integrators: the energy is oscillating, displaying multiple different frequencies, but remains bounded nevertheless. On the contrary, the behaviour of the total energy for the non-variational enforced orbit-averaging scheme completely blows up as seen in Figure~\ref{fig:long_time_nonvar}.

%% file: summary.tex
\section{Summary}\label{sec:summary}
In this paper, we have introduced two possible subcycling strategies for variational GEMPIC methods addressing the Vlasov--Maxwell system in magnetized plasmas. The first one is a straightforward upgrade of the existing variational GEMPIC methods, specifically of the ones discussed in Ref.~\cite{Squire-Qin-Tang-PIC:2012PhPl} and Ref.~\cite{Xiao-et-al:2018PlST}. The algorithm was tested both in electrostatically and electromagnetically dominated cases. The tests revealed that the resulting, rather peculiar subcycling scheme---magnetic field is properly orbit-averaged but the electric-field impulse evaluated only once per the subcycling period---may result in artificial oscillations if the electric field impulse is too strong in relation to the magnetic field impulse. The root cause was verified by enforcing the electric-field orbit-averaging which removed the spurious oscillations but would result in a non-variational particle push. We have performed also low-resolution 3-D simulations and the results remain qualitatively the same.

Our second strategy is aimed at mitigating the possible limitations of the first algorithm. Instead of relying on the ``summation-by-parts'' trick, which is the corner stone of the existing electromagnetically gauge-invariant variational GEMPIC methods, we considered the possibility of performing genuine integration by parts instead. This lead us to suggest an algorithm where the orbit-averaging is done properly for both the electric and magnetic impulse and which retains the gauge invariance and hence the algebraic charge-conservation law. Numerical tests confirmed our hypothesis and the artificial oscillations completely vanished. The trade-off with the second algorithm is that it requires a global implicit solve for the electric field, the current density, and the particle push are entangled. Furthermore, it appears to be difficult to find an explicit scheme that would handle the electric and magnetic fields equally. It seems to be necessary to treat the electromagnetic potential as being time-continuous for the sake of performing partial integrations in the field--particle-interaction part of the discrete action and this tightly couples the degrees of freedom for the fields to the degrees of freedom for the particles during the synchronizing global steps, effectively resulting in a globally implicit scheme. The non-synchronous particle steps fortunately remain decoupled and lead to only individually implicit push. 

While the first approach may introduce a visible error, the overal macroscopic behaviour of the algorithm in our tests nevertheless appeared to be reasonable, even when the global time step was pushed beyond the cyclotron period. Since this particular strategy also admits a fully explicit scheme, including the subcycling of the particle orbits, it would be interesting to test whether the introduced error---in our tests the error remains acceptable when using a moderate number of substeps and a global step size below the cyclotron period---would remain acceptable in large-scale simulations such as the ones performed in Ref.~\cite{Xiao-Qin-6d-tokamak:2020arXiv}. If the error would not turn out to be prohibitively large, the fully explicit scheme with subcycling could have the potential to reduce the complexity of such simulations significantly.

Overall, it remains to be seen whether subcycling of particle orbits in explicit, variational geometric particle-in-cell methods with proper orbit-averaging of both electric and magnetic impulses is possible. Since our fully implicit scheme displays all the desired properties---apart from being implicit---a fully explicit variational scheme with similar orbit-averaging would likely become the goto method among the plethora of particle-in-cell algorithms. In the future, we aim to investigate also the possibility of adaptive temporal integration. Such algorithm could be ideal from the perspectives that the guiding magnetic field may vary spatially. 

